\documentstyle[preprint,epsbox,seceq]{jpsj}
\def\alt{\mathrel {\mathpalette \vereq <}}
\def\agt{\mathrel {\mathpalette \vereq >}}
\def\mathpalette#1#2{\mathchoice {#1\displaystyle {#2}}%
                                 {#1\textstyle {#2}}%
                                 {#1\scriptstyle {#2}}%
                                 {#1\scriptscriptstyle
{#2}}}
\def\vereq#1#2{\lower 3pt\vbox {\baselineskip 1.5pt \lineskip 1.5pt \ialign
{$\m@th #1\hfill ##\hfil $\crcr #2\crcr \sim \crcr }}}
\def\m@th{\mathsurround=0pt}

\def\runtitle{
Electrical Conductivity of Fermi Liquids. II. 
}
\def\runauthor{Takuya OKABE}

\title{
Electrical Conductivity of Fermi Liquids. II. \\
Quasiparticle Transport
}

\author
{ 
Takuya {\sc Okabe}\footnote{Present address: 
Faculty of Engineering, Gunma University, Kiryu, Gunma 376.
E-mail: okabe@phys.eg.gunma-u.ac.jp}
}
\inst
{
Department of Physics, Kyoto University, Kyoto 606-01
}

\recdate
{
March 12, 1998
}

\abst
{
We develop a general theory of Fermi liquids to 
discuss the Kadowaki-Woods relation $A\propto \gamma^2$
between the coefficient $A$ of 
the $T^2$ term in the resistivity $\sigma^{-1}=AT^2$
and $\gamma$ of the specific heat $C=\gamma T$
at low temperatures.
We derive a formula for the ratio $A/\gamma^2$
which is expressed as
a product of two dimensionless parameters $\alpha$ and $F$,
where $\alpha$
represents a coupling constant for quasiparticle scattering 
and $F$ is a geometric factor 
determined by the shape of the Fermi surface.
Then we argue that the universal ratio $A/\gamma^2=
1.0\times 10^{-5} \mu {\Omega}\mbox{cm}(\mbox{mol}\,\mbox{K/mJ})^2$
observed in heavy fermion compounds is reproduced 
under the conditions $\alpha\sim 1$ and $F\sim 20$. 
The former is regarded as 
a universality of Fermi liquids in a strong coupling regime,
and the latter is corroborated by
evaluating $F$ definitely in simple cases.
It is noted that the proportional relation is just 
an example of the universal phenomena to be expected for
the whole class of strong coupling Fermi liquids.
}

\kword
{Fermi Liquid Theory, electrical conductivity,
Boltzmann equation, many-body effect,  quasiparticle transport
}

\begin{document}
\sloppy
\maketitle
\section{Introduction}

In the previous paper,
we discussed the many-body effect on the Drude weight 
on the basis of the Fermi liquid theory,~\cite{rf:I}\
and showed that the Drude weight
is directly related
not to the velocity $v^*_k$ but 
to the current $j^*_k$ of quasiparticle.
It was stressed that 
the effective mass defined for $j^*_k$
is generally different from the thermal mass
estimated from the $T$ linear coefficient of the specific heat.
Although we may work with the zero temperature formalism 
to discuss the Drude weight and the other phenomena 
in a collisionless regime,
the effect of quasiparticle damping should be taken into account 
in a hydrodynamic regime, for example,
to investigate dc conductivity.
In this article, we develop a general theory 
to discuss both regimes in an unified manner
and derive a formula for the Kadowaki-Woods ratio
$A/ \gamma^2$ between the coefficient $A$ of 
the $T^2$ term in the resistivity $\sigma^{-1}=AT^2$
and $\gamma$ of the specific heat $C=\gamma T$.
Then we discuss the conditions for 
the Kadowaki-Woods relation $A\propto \gamma^2$
and estimate an universal value taken by the ratio $A/ \gamma^2$.

A framework for this purpose was first provided by
\'Eliashberg.~\cite{rf:Eliashberg} \
His formalism is sophisticated and firmly established
so that we take this as a starting point 
to discuss electrical properties of Fermi liquids.
Therefore a large part of this paper is devoted to 
recapitulation of his formalism.
This is necessary not only to introduce notations, but
because \'Eliashberg mainly concentrated upon 
a microscopic theory of zero sound,~\cite{rf:Eliashberg,rf:Eliashberg2}
while we are interested in a general theory 
of quasiparticle transport, 
especially the electrical conductivity.
Thus in the first half of this paper
we derive a linearized transport equation
in a general but concise form,
so as to serve our purpose of the latter sections.
All the results will be expressed in terms of 
renormalized quantities characterizing quasiparticles,
and then we discuss transport properties of Fermi liquids 
in general terms.

For heavy fermion compounds at low temperatures, 
it has been known that the ratio $A/\gamma^2$ takes a universal value
\(
\simeq 1.0
\times 10^{-5} \mu {\Omega}\mbox{cm}(\mbox{mol}\,\mbox{K/mJ})^2.
\)~\cite{rf:KW}\
In view of this experimental fact, we must 
theoretically explain not only 
(i) the reason why the proportional relation $A\propto \gamma^2$
holds for a certain class of systems,
but also (ii) the value of the universal ratio itself.
These points have been
discussed theoretically 
by many authors.~\cite{rf:YY,rf:MMV,rf:KWrel1,rf:KWrel2}\
However, the theories are more or less 
based on some specific models or approximations.
Among them, 
the theories presented in refs.~\citen{rf:MMV,rf:KWrel1,rf:KWrel2}
are not enough for us to study 
the point (ii) microscopically,
as their results are indefinite up to a factor of order unity 
related to effectiveness of Umklapp processes 
responsible for a finite conductivity.
Therefore we shall follow the microscopic theory
of Yamada and Yosida~\cite{rf:YY}
developed on the basis of the periodic Anderson model,
in which the effect of Umklapp processes is explicitly specified.
Unlike ref.~\citen{rf:YY}, however, 
we try to keep the theory as general as possible. 
This is because we believe that 
such a universal relation
as observed by Kadowaki and Woods~\cite{rf:KW} should be explained 
within the context of general theory of Fermi liquids
without making any specific assumptions.
Moreover we take a method different from ref.~\citen{rf:YY}
to derive a formula for $A/\gamma^2$, thereby
the fully renormalized three-point vertex function
$\tilde{\mib{{\Lambda}}}_p$ introduced by Yamada and Yosida
is shown to be related to
the deviation from local equilibrium
quasiparticle distribution~\cite{rf:PN}\
${\varphi}_{p}\propto \delta\bar{n}_p^0$,
(eqs.~(\ref{defdelbarn}) and (\ref{varphiYY})).
Once we obtain a definite general expression for $A/\gamma^2$,
the observed universality suggests that it would give 
a reasonable estimate for the ratio
even if it is applied for a simple model. 
As a matter of fact,
we deal with some simple models to evaluate the ratio.
At the outset, however, 
no assumption is made except for those assuring the applicability of 
the Fermi liquid theory,  which are briefly sketched below.

We discuss a single-band electron system
with the band dispersion $\varepsilon_p$ for the momentum $p$
and the velocity $v_p=\partial \varepsilon_p/\partial p$.
We do not assume an explicit form of electron-electron interaction,
except that it must conform to the basic 
hypothesis of the Fermi liquid theory:
We assume a well-defined quasiparticle $p$
of the energy dispersion $\varepsilon^*_p$
which is formally obtained from 
non-interacting state $\varepsilon_p$ by perturbation theory.
The interaction between quasiparticles 
is described by the scattering amplitude $A_{pp'k}=A(p,p';p'+k,p-k)$
which also, in principle, 
is to be derived from a microscopic model.
At absolute zero, 
$A(p_1,p_2;p_3,p_4)$ 
is assumed to be a real and regular 
function of $p_i$ ($i=1\sim 4$)
around the Fermi surface, $\varepsilon^*_{p_i}=\mu$.~\cite{rf:com0}\
As we discuss a linear response 
of the system to a macroscopic plane-wave perturbation
with a sufficiently small wave number $k$ and frequency $\omega$,
the energy dependence of $A_{pp'k}$
gives rise only to correction terms of higher order than we need.
Thus the energy variables of $A_{pp'k}$ are
fixed to take the Fermi energy $\mu$.
The procedure we follow
is a perturbation theory with respect to $k$ and $\omega$,
i.e., at $T=0$ only the terms linear in $k$ and $\omega$ will be
retained; 
still the effect of electron-electron interaction is
taken into account to an arbitrary order.
At finite temperature, 
the quasiparticle scattering $A_{pp'k}$
induces a finite decay rate $\gamma^*_p(T)$ 
for the quasiparticle $p$.
The imaginary part of the scattering amplitude 
is found to be the same order as $\gamma^*_p(T) \sim T^2$.
Under the condition 
$|\varepsilon_p^* -\mu|\gg \gamma^*_p$,
the quasiparticle peak in the one-particle Green's function 
$G(p,\varepsilon)$ is approximated by the delta function 
$\delta(\varepsilon-\varepsilon_p^*+\mu)$
and the quasiparticle $p$ is well defined.
Therefore, we have to be in a low temperature region
$T \alt T_0$, where $T_0$ is roughly estimated by 
$|\varepsilon_p^*-\mu|\sim T \simeq \gamma^*_p(T_0)$.
We neglect the terms of order $O(\gamma^{*\,2}_p)$.
To sum up, we regard $k$, $\omega$ and $\gamma^*_p$
as the small parameters with which 
to develop a perturbation expansion
of the electrical conductivity $\sigma(\mib{k},\omega)$,
where the singular terms due to the quasiparticle poles
are retained systematically.

In the next section, 
we follow \'Eliashberg~\cite{rf:Eliashberg} 
to transform the Kubo formula of
the electrical conductivity $\sigma_{\mu\nu}(\mib{k},\omega)$,
expressed by temperature Green's functions and vertex functions,
into a form suitable for our purpose.
In \S3, we investigate an explicit form of the vertex functions
to derive a set of formulas to 
calculate $\sigma_{\mu\nu}(\mib{k},\omega)$.
One of our results is microscopic derivation of 
a linearized Boltzmann equation.
In \S3.1, the collisionless regime $\omega\gg\gamma^*_p$
is investigated, where, after elucidating 
the quantities defined in \S2 in physical terms,
we derive the formula for the optical conductivity 
obtained in the previous paper.~\cite{rf:I}\
In \S3.2, 
the hydrodynamic regime $\omega\ll\gamma^*_p$ is investigated,
where we derive the collision integral of the Boltzmann equation
using results of Yamada and Yosida,~\cite{rf:YY}\ 
and estimate the low frequency limit
of the incoherent part of the conductivity due to
quasiparticle scattering.
Using the result of \S3.2,
we derive a formula for the coefficient $A=\sigma(0)^{-1}/T^2$.
Then in \S4 we discuss the universality 
of the Kadowaki-Woods relation $A\propto \gamma^2$.
With the results obtained for Fermi surfaces of simple examples,
we estimate the ratio $A/\gamma^2$ and find
that the observed ratio is reasonably
reproduced for a strong coupling Fermi liquid.
Discussion and conclusion are presented in \S\S5 and 6 respectively.

\section{\'Eliashberg Formalism}
According to the linear response theory, 
the electrical conductivity $\sigma_{\mu\nu}(\mib{k},\omega)$
is given by 
\fulltext   
\begin{equation}
\sigma_{\mu\nu}(\mib{k},\omega)
=
\frac{e^2}{\Omega}\sum_{p,p'}
v_{p\mu}
\frac{K^R_{pp'}(\mib{k},\omega)-K^R_{pp'}(\mib{k},0)}{\mbox{i}\omega}
v_{p'\nu},
\label{sigma(k,omega)}
\end{equation}
where $\Omega$ is the volume of the system, and 
the function $K^R_{pp'}(\mib{k},\omega)$ is 
defined by 
\begin{equation}
K^R_{pp'}(\mib{k},\mbox{i}\omega_m)=
K_{pp'}(\mib{k},\omega_m)\quad {\rm for}\quad \omega_m>0,
\end{equation}
\begin{equation}
K_{pp'}(\mib{k},\omega_m)=\frac{1}{2}\int^{1/T}_{-1/T}
\mbox{e}^{\omega_m\tau}
K_{pp'}(\mib{k},\tau)\mbox{d}\tau,
\end{equation}
\begin{equation}
K_{pp'}(\mib{k},\tau)=
\langle T_\tau \mbox{e}^{(H-\mu N)\tau}
\hat{c}^\dagger_{p'-k/2}
\hat{c}_{p'+k/2}
\mbox{e}^{-(H-\mu N)\tau}
\hat{c}^\dagger_{p+k/2}
\hat{c}_{p-k/2}\rangle.
\end{equation}
After the analytic continuation with respect to $\omega$,
one finds~\cite{rf:Eliashberg}\
\begin{eqnarray}
K^R(k)&=&-\frac{1}{4\pi\mbox{i}}\int^\infty_{-\infty}
\mbox{d}\varepsilon
\left[
\tanh\frac{\varepsilon-\omega/2}{2T}K_1(p,k)
+\left(\tanh\frac{\varepsilon+\omega/2}{2T}
-\tanh\frac{\varepsilon-\omega/2}{2T}\right)K_2(p,k)
\right.\nonumber\\
&&\left.
-\tanh\frac{\varepsilon+\omega/2}{2T}K_3(p,k)
\right],
\end{eqnarray}
where
\begin{equation}
K_i(p,k)=g_i(p,k)
\left[
1+\frac{1}{4\pi\mbox i}
\int^\infty_{-\infty}
\mbox{d}\varepsilon'
\sum_{k=1}^3
T_{ik}(p,p';k)
g_k(p',k)\right],
\label{K=g(1+Tg)}
\end{equation}
\halftext
and
\begin{eqnarray}
g_1(p,k)
&=&G^{R}(p+k/2)G^{R}(p-k/2),\nonumber\\
g_2(p,k)
&=&G^{R}(p+k/2)G^{A}(p-k/2),\nonumber\\
g_3(p,k)
&=&G^{A}(p+k/2)G^{A}(p-k/2).\label{gi}
\end{eqnarray}
Hereafter we use the abbreviated notation such as 
\[p=(\mib{p},\varepsilon),\quad
p'=(\mib{p}',\varepsilon'),\quad k=(\mib{k},\omega).\]
The retarded Green's function $G^R(p)$ is given by
\begin{equation}
G^R(p)=\frac{1}{\varepsilon-\varepsilon_p+\mu
-\Sigma^R(\mib{p},\varepsilon)},
\end{equation}
where $\Sigma^R(\mib{p},\varepsilon)$ 
is the retarded selfenergy.

\begin{figure}[t]
\centerline{\epsfile{file=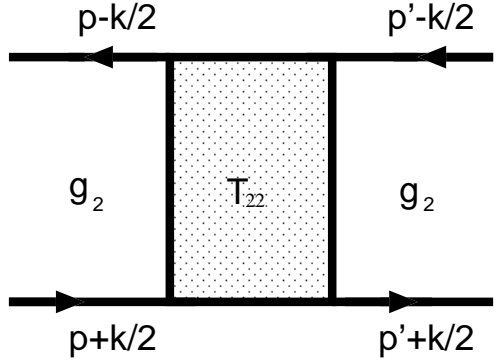,width=6.6cm}}
\caption{Diagrammatic expressions for 
$g_2(p,k)$, $T_{22}(p,p';k)$ and $g_2(p',k)$
(from left to right).
}
\label{fig:T22}
\end{figure}

In eq.~(\ref{K=g(1+Tg)}), $T_{ij}(p,p';k)$  for $i,j =1,2,3$
represent properly antisymmetrized vertex functions.
Among them, $T_{22}(p,p';k)$,
shown in Fig.~\ref{fig:T22} together 
with $g_2(p,k)$ of eq.~(\ref{gi}),
plays an important role
to determine quasiparticle transport properties.
The function $T_{22}(p,p';k)$ 
is expressed in terms of $\Gamma(p,p';k)$,
the four-point vertex function analytic in the region defined by
\begin{eqnarray}
\mbox{Im}(\varepsilon-\omega/2)<0,&\quad &
\mbox{Im}(\varepsilon+\omega/2)>0,\nonumber\\
\mbox{Im}(\varepsilon'-\omega/2)<0,&\quad &
\mbox{Im}(\varepsilon'+\omega/2)>0.
\end{eqnarray}
In fact, 
this region must be further divided into several portions
in accordance with the analytic property of $\Gamma(p,p';k)$.
As a result, we obtain~\cite{rf:Eliashberg}\
\fulltext
\begin{eqnarray}
T_{22}(p,p';k)
&=&\left(
\tanh\frac{\varepsilon'+\omega/2}{2T}
-\tanh\frac{\varepsilon'-\omega/2}{2T}
\right)\mbox{Re}\,
\Gamma(p,p';k)
\nonumber\\
&&+\coth\frac{\varepsilon'-\varepsilon}{2T}
\left(
\Gamma^{\rm II}(p,p';k)
-
\Gamma^{\rm III}(p,p';k)
\right)
-\tanh\frac{\varepsilon'-\omega/2}{2T}
\mbox{Im}\,
\Gamma^{\rm II}(p,p';k) \nonumber\\
&&
+\coth\frac{\varepsilon'+\varepsilon}{2T}
\left(
\Gamma^{\rm III}(p,p';k)
-
\Gamma^{\rm IV}(p,p';k)
\right)
+\tanh\frac{\varepsilon'+\omega/2}{2T}\mbox{Im}\,
\Gamma^{\rm IV}(p,p';k),
\label{T22=mendo}
\end{eqnarray}
\halftext
where $\Gamma^{\rm II}(p,p';k)$,
$\Gamma^{\rm III}(p,p';k)$ and $\Gamma^{\rm VI}(p,p';k)$
are analytic respectively in
\begin{eqnarray}
{\rm II}&:& \mbox{Im}(\varepsilon-\varepsilon')<0, 
\quad \mbox{Im}(\varepsilon+ \varepsilon')>0, \nonumber\\
{\rm III}&:& \mbox{Im}(\varepsilon-\varepsilon')>0, 
\quad \mbox{Im}(\varepsilon+ \varepsilon')>0, \nonumber\\
{\rm IV}&:& \mbox{Im}(\varepsilon-\varepsilon')>0, 
\quad \mbox{Im}(\varepsilon+ \varepsilon')<0. 
\label{II&III&IV}
\end{eqnarray}
The imaginary part of $\Gamma(p,p';k)$ has 
discontinuity at the boundary of the regions
defined in eq.~(\ref{II&III&IV}),
so that the imaginary parts of 
$\Gamma^{\rm II}(p,p';k)$,
$\Gamma^{\rm III}(p,p';k)$ and $\Gamma^{\rm IV}(p,p';k)$
are generally different,
while the real part of them takes a common value 
$\mbox{Re}\Gamma(p,p';k)$.
Accordingly,  we wrote eq.~(\ref{T22=mendo})
in such a way that 
the first line in the right-hand side is real 
but the second and third lines are pure imaginary.
As for the functions $T_{ij}$ other than $T_{22}$, 
it is enough for us to note
\begin{eqnarray}
T_{12},T_{32},T_{21},T_{23}
&\propto&
\tanh\frac{\varepsilon'+\omega/2}{2T}
-\tanh\frac{\varepsilon'-\omega/2}{2T}\nonumber\\
&=&O(\omega).
\quad(\omega\rightarrow 0)
\label{Ti2}
\end{eqnarray}

The energy $\varepsilon_p^*$ and the decay rate $\gamma^*_p$
of the quasiparticle $p$ are defined by 
the real and imaginary part of 
\begin{equation}
G^R(\mib{p},\varepsilon_p^*-\mu-\mbox{i}\gamma^*_p)^{-1}=0,
\end{equation}
namely, by
\begin{equation}
\varepsilon_p^*-\mu=z_p(\varepsilon_p+\mbox{Re}\Sigma(\mib{p},0)-\mu),
\label{varepsilon*}
\end{equation}
and
\begin{equation}
\gamma^*_p=-z_p\mbox{Im}\,
\Sigma^R(\mib{p},0),
\label{gamma*p}
\end{equation}
where the renormalization factor $z_p$ is given by
\begin{equation}
z_p^{-1}=\left.
1-\frac{\partial}{\partial \varepsilon}
\mbox{Re}\,\Sigma^R(\mib{p},\varepsilon)
\right|_{\varepsilon=0}.
\end{equation}
For $\varepsilon\sim |\varepsilon_p^*-\mu|\gg \gamma^*_p$,  
we may write
\begin{eqnarray}
\mbox{Im}\,G^R(p)&=&-\pi z_p \delta(\varepsilon-
\varepsilon^*_p+\mu),
\label{ImGR=}
\end{eqnarray}
and 
\begin{equation}
g_2(p,k)
=\frac{2\pi\mbox{i}z_p^2
\delta(\varepsilon-\varepsilon^*_p+\mu)}
{\omega-\mib{v^*_p\cdot k}+2\mbox{i}\gamma^*_p},
\label{g2=}
\end{equation}
where
\begin{equation}
v^*_{p\mu}=\frac{\partial \varepsilon_p^*}{\partial p_\mu}.
\end{equation}
These results are legitimately used
in the following discussion, as we shall find that
the energy region relevant to us is
$\varepsilon\sim |\varepsilon_p^*-\mu|\sim T \gg \gamma^*_p$.
Equation~(\ref{g2=}) is valid as far as 
the linear terms with respect to $\omega$, $k$ and $\gamma_p^*$ 
are concerned.
In effect, an essential singularity 
in the linear response to a macroscopic perturbation
is brought about by 
the quasiparticle-quasihole pair propagator $g_2(p,p';k)$.
The function $g_1(p,p';k)$ and $g_3(p,p';k)$ 
are less singular than  $g_2(p,p';k)$, and 
these functions, combined with $T_{ij}$ ($i,j=1$ or 3),
give rise only to the vertex correction,
$v_{p\mu}\rightarrow \Lambda_{p\mu}$ in eq.~(\ref{sigma(k,omega)}).
As a result, the conductivity of Fermi liquids is given by
\fulltext
\begin{eqnarray}
\sigma_{\mu\nu}(k,\omega)
&=&\frac{e^2}{2\Omega}
\int\frac{\mbox{d}^4 p}{(2\pi)^4}
\frac{1}{2T}\cosh^{-2}\frac{\varepsilon}{2T}
\Lambda_{p\mu}g_2(p,k)\left(\Lambda_{p\nu}-
\frac{\mbox i}{2}
\int\frac{\mbox{d}^4 p'}{(2\pi)^4}
T_{22}(p,p';k)g_2(p',k)
\Lambda_{p'\nu}\right).
\label{sigma2}
\end{eqnarray}
\halftext
In fact,
by the close inspection of \'Eliashberg's derivation~\cite{rf:Eliashberg},
we can verify that the relation,
\begin{equation}
z_p\Lambda_{p\mu}=v^*_{p\mu},
\label{zpLambda}
\end{equation}
holds for the vertex function \(\Lambda_{p\mu}\)
under the assumption of the Fermi liquid theory
to neglect the imaginary part of $\Gamma(p,p';k)$.~\cite{rf:com}\

Noting
\begin{equation}
\frac{1}{4T}\cosh^{-2}\frac{\varepsilon^*_p-\mu}{2T}
=
-\frac{\partial n^0_p}{\partial \varepsilon^*_p},
\label{cosh-2}
\end{equation}
\begin{equation}
n^0_p=n^0(\varepsilon^*_p)=
\frac{1}{\mbox{e}^{(\varepsilon^*_p-\mu)/T}+1},
\end{equation}
and using eqs.~(\ref{sigma2}) and  (\ref{zpLambda}),
we write the total current $j_\mu(k,\omega)$ per volume 
as
\begin{eqnarray}
j_\mu(k,\omega)&=&\sum_{\nu}
\sigma_{\mu\nu}(k,\omega)E_\nu
=
\frac{e}{\Omega}\sum_p
\delta \bar{n}_p v^*_{p\mu},
\label{sigma}
\end{eqnarray}
where
\begin{eqnarray}
\delta\bar{n}_p &\equiv&
\frac{\partial n^0_p}{\partial \varepsilon^*_p}{\varphi}_p,
\label{delbarn}
\end{eqnarray}
and
\fulltext
\begin{eqnarray}
{\varphi}_p=-e\int^\infty_{-\infty}\frac{\mbox{d}\varepsilon}{2\pi}
z_p^{-1}g_2(p,k)\left(\mib{\Lambda_{p}\cdot E}-
\frac{\mbox i}{2}
\int\frac{\mbox{d}^4 p'}{(2\pi)^4}
T_{22}(p,p';k)g_2(p',k)
\mib{\Lambda_{p'}\cdot E}\right).
\label{varphi0}
\end{eqnarray}
\halftext
The function $T_{22}(p,p';k)$, by definition,
is the vertex part
which is connected to $g_2(p,k)$ on the left end
and to $g_2(p',k)$ on the right end (Fig.~\ref{fig:T22}).
Thus, in general, 
$T_{22}(p,p';k)$ may comprise the sector $g_i$ $(i \ne 2)$
and $T_{ij}$ ($i, j\ne 2$) in between 
the $g_2$ functions on both ends.
However,  because of eq.~(\ref{Ti2}),
we may neglect the portions $T_{ij}$ ($i, j\ne 2$) 
which may appear in $T_{22}$
to the accuracy of order $O(\omega)$.
Hence, for our purpose, it is convenient 
to introduce the vertex $T^{(0)}_{22}(p,p';k)$
which is irreducible with respect to the section $g_2$;
\fulltext
\begin{equation}
T_{22}(p,p';k)=T^{(0)}_{22}(p,p';k)
-\frac{\mbox i}{2}\int\frac{\mbox{d}^4 p''}{(2\pi)^4}
T^{(0)}_{22}(p,p'';k)g_2(p'',k)T_{22}(p'',p';k),
\label{T22=T022-}
\end{equation}
Then, in terms of $T^{(0)}_{22}(p,p';k)$, 
eq.~(\ref{varphi0}) is cast into 
\begin{eqnarray}
{\varphi}_p
=-e\int^\infty_{-\infty}\frac{\mbox{d}\varepsilon}{2\pi}
z_p^{-1}g_2(p,k)\mib{\Lambda_{p}\cdot E}
-\frac{\mbox i}{2}
\int^\infty_{-\infty}\frac{\mbox{d}\varepsilon}{2\pi}
\int\frac{\mbox{d}^4 p'}{(2\pi)^4}
z_p^{-1}g_2(p,k)T^{(0)}_{22}(p,p';k)z_{p'}\phi_{p'},
\label{varphip=..}
\end{eqnarray}
\halftext
where $\phi_p$ is defined by
\begin{equation}
{\varphi}_p=\int^\infty_{-\infty}\frac{\mbox{d}\varepsilon}{2\pi}
\phi_p.
\end{equation}
As the function $\varphi_{p}$ is always multiplied by 
\({\partial n^0_p}/{\partial \varepsilon^*_p},\)
we may use eq.~(\ref{g2=}) in eq.~(\ref{varphip=..}).
Noting that
\[\phi_p \propto g_2(p,k)\propto
\delta(\varepsilon-\varepsilon^*_p+\mu),
\]
we may put
\[{\phi}_p =
2\pi\delta(\varepsilon-\varepsilon^*_p+\mu)\varphi_p.
\]
Therefore, from eq.~(\ref{varphip=..}) with 
eq.~(\ref{g2=}), we obtain~\cite{rf:Eliashberg}
\fulltext
\begin{eqnarray}
\mbox{i}(\omega-\mib{v^*_p\cdot k}){\varphi}_p&=&
e\mib{v^*_p\cdot E}
+\frac{\mbox{i}}{2}
\int\frac{\mbox{d}^4 p'}{(2\pi)^3}
\delta(\varepsilon'-\varepsilon^*_{p'}+\mu)
z_p z_{p'}
\left.
T^{(0)}(p,p';k)\right|_{\varepsilon=0}{\varphi}_{p'}
+2\gamma^*_p\varphi_p,
\label{bolprttype}
\end{eqnarray}
for the momentum $p$ on the Fermi surface, 
$\varepsilon^*_{p}=\mu$.
As the vertex functions 
which appear in the following section are 
all of the type $T_{22}^{(0)}(p,p';k)$,
this is simply written as $T^{(0)}(p,p';k)$ 
without the subscripts.

\section{Four-Point Vertex Function}
\subsection{Real part of $T^{(0)}$: 
Collisionless Regime, $\omega\gg \gamma^*_p$}
The four-point vertex function $T^{(0)}$
in eq.~(\ref{bolprttype}) 
is divided  into a real and imaginary part.
Since eq.~(\ref{T22=mendo}) for $T$ and $\Gamma$ 
can be used as well for $T^{(0)}$ and $\Gamma^{(0)}$,
using the first line of eq.~(\ref{T22=mendo})
and eq.~(\ref{cosh-2}), we obtain
\fulltext
\begin{eqnarray}
\lefteqn{\frac{\mbox{i}}{2}
\int\frac{\mbox{d}^4 p'}{(2\pi)^3}
\delta(\varepsilon'-\varepsilon^*_{p'}+\mu)
z_p z_{p'}
\left.
T^{(0)}(p,p';k)\right|_{\varepsilon=0}{\varphi}_{p'}}\\
&& \quad =-
{\mbox{i}\omega}
\sum_{p'}
A(\mib{p},\mib{p'};\mib{k})
\frac{\partial n^0}{\partial \varepsilon^*_{p'}}{\varphi}_{p'}
-\frac{1}{2}
\int\frac{\mbox{d}^4 p'}{(2\pi)^3}
\delta(\varepsilon'-\varepsilon^*_{p'}+\mu)
z_p z_{p'}
\left.
\mbox{Im}\,T^{(0)}(p,p';k)\right|_{\varepsilon=0}{\varphi}_{p'},
\label{Re+ImT}
\end{eqnarray}
\halftext
where the function $A(\mib{p},\mib{p'};\mib{k})$ signifies
the scattering amplitude of quasiparticles on the Fermi surface,
\begin{equation}
A(\mib{p},\mib{p'};\mib{k})\equiv
z_pz_{p'}
\left.\mbox{Re}\,\Gamma^{(0)}(p,p';k)
\right|_{\scriptstyle{\varepsilon=\varepsilon'=0\atop \omega=0}}.
\label{App'k}
\end{equation}
To the accuracy of our concern,
we may put $k=0$ in the first term of the right-hand side
of eq.~(\ref{Re+ImT}).
Consequently, in terms of the forward scattering amplitude,
\begin{equation}
A(\mib{p},\mib{p'})\equiv
A(\mib{p},\mib{p'};\mib{k}=0),
\label{App'}
\end{equation}
eq.~(\ref{bolprttype}) is written
\begin{equation}
-\mbox{i}\omega\psi_p
+\mbox{i}\mib{v^*_p\cdot k}{\varphi}_p
+e\mib{v^*_p\cdot E}
=I({\varphi}),
\label{boltz}
\end{equation}
where
\begin{equation}
\psi_p=
{\varphi}_p+
\sum_{p'}
A(\mib{p},\mib{p'})
\frac{\partial n^0}{\partial \varepsilon^*_{p'}}{\varphi}_{p'},
\label{psi}
\end{equation}
and 
\fulltext
\begin{equation}
I({\varphi})=
\frac{1}{2}
\int\frac{\mbox{d}^4 p'}{(2\pi)^3}
\delta(\varepsilon'-\varepsilon^*_{p'}+\mu)
z_p z_{p'}
\left.
\mbox{Im}\,T^{(0)}(p,p';k)\right|_{\varepsilon=0}{\varphi}_{p'}
-2\gamma^*_p{\varphi}_p.
\label{I(varphi)=}
\end{equation}
\halftext
In the following part of this subsection, 
we discuss the left-hand side of eq.~(\ref{boltz}),
neglecting $I({\varphi})$ of the right-hand side. 
As we shall see, this is justified when $\omega\gg \gamma^*_p$,
i.e., in the collisionless regime.

To see the physical meaning of $\psi_p$ defined by eq.~(\ref{psi}),
let us define
\[
f(\mib{p},\mib{p'})=z_pz_{p'}
\left.\mbox{Re}\,\Gamma(p,p';k)
\right|_{{k}=0\atop \omega\rightarrow 0}.
\]
Then, using 
\begin{equation}
g_2(p,k)\simeq
\frac{2\pi\mbox{i}z_p^2}{\omega} \delta(\varepsilon-\varepsilon^*_p+\mu),
\quad (\mib{k}=0,\, \omega\gg\gamma_p^*)
\label{g2pk}
\end{equation}
from eq.~(\ref{T22=T022-}), we obtain
\fulltext
\begin{equation}
f(\mib{p},\mib{p'})=A(\mib{p},\mib{p'})
-
\sum_{p''}
A(\mib{p},\mib{p''})
\frac{\partial n^0}{\partial \varepsilon^*_{p''}} 
f(\mib{p''},\mib{p'}),
\qquad (\omega\gg \gamma^*_p)
\label{f=A-Af}
\end{equation}
\halftext
Therefore the function $f(\mib{p},\mib{p'})$,
related with $A(\mib{p},\mib{p'})$ by eq.~(\ref{f=A-Af}),
is identified as the Landau function of the Fermi liquid theory.
In particular, in an isotropic system, 
$f(\mib{p},\mib{p'})$ and 
$A(\mib{p},\mib{p'})$
are functions of $\cos \theta=\mib{p\cdot p'}/p^2_{\rm F}$,
so that they may be expanded in spherical harmonics,
\begin{eqnarray}
A(\mib{p}\sigma,\mib{p'}\sigma')
&=&\frac{\pi^2 \hbar^3}{\Omega m^* p_{\rm F}}
\sum_l (A_l^s+\mib{\sigma\cdot\sigma'} A_l^a)
P_l(\cos\theta),\nonumber\\
f(\mib{p}\sigma,\mib{p'}\sigma')
&=&\frac{\pi^2 \hbar^3}{\Omega m^* p_{\rm F}}
\sum_l (F_l^s+\mib{\sigma\cdot\sigma'} F_l^a)
P_l(\cos\theta),\nonumber
\end{eqnarray}
where
\begin{equation}
\frac{ m^* p_{\rm F}}{2\pi^2 \hbar^3}=
-\left.\frac{\partial n_p^0}{\partial \varepsilon^*_{p}} 
\right|_{p=p_{\rm F}}.
\end{equation}
Here we have explicitly introduced the spin indices.
Then, eq.~(\ref{f=A-Af}) gives
\begin{equation}
\frac{A_l^s}{2l+1}=
\dfrac{\dfrac{F_l^s}{2l+1}}{1+\dfrac{F_l^s}{2l+1}},
\qquad
\frac{A_l^a}{2l+1}=\dfrac{\dfrac{F_l^a}{2l+1}}{1+\dfrac{F_l^a}{2l+1}}.
\label{A_l^s=F}
\end{equation}

For $\delta n_p$ defined by
\begin{equation}
\delta n_p \equiv 
\frac{\partial n^0_p}{\partial \varepsilon^*_p}\psi_p,
\label{deln}
\end{equation}
and $\delta \bar{n}_p$, eq.~(\ref{delbarn}),
there are relations 
\begin{equation}
\delta n_p=\delta \bar{n}_p+
\frac{\partial n^0_p}{\partial \varepsilon^*_p}
\sum_{p'}
A(\mib{p},\mib{p'})
\delta \bar{n}_{p'},
\label{deln=delbarn+}
\end{equation}
and
\begin{equation}
\delta \bar{n}_p=\delta {n}_p-
\frac{\partial n^0_p}{\partial \varepsilon^*_p}
\sum_{p'}
f(\mib{p},\mib{p'})
\delta {n}_{p'},
\label{delnbar=deln}
\end{equation}
which are obtained by using eqs.~(\ref{psi}) and (\ref{f=A-Af}).
Equation~(\ref{delnbar=deln}) suggests
the following relations,
\begin{eqnarray}
\delta \bar{n}_p
&\equiv&n_p-\bar{n}_p^0, 
\quad
\delta {n}_p
\equiv n_p-{n}_p^0, \label{defdelbarn}\\
\bar{n}_p^0&\equiv&n^0(\bar{\varepsilon}^*_p),
\quad
{n}_p^0 \equiv n^0({\varepsilon}^*_p),\\
\bar{\varepsilon}^*_p&\equiv&
\varepsilon^*_p+
\sum_{p'}
f(\mib{p},\mib{p'})\delta {n}_{p'},
\end{eqnarray}
Therefore, $\delta \bar{n}_p$ is regarded as the deviation 
from $\bar{n}_p^0$, the equilibrium distribution function 
for the local excitation energy $\bar{\varepsilon}^*_p$
and $\delta {n}_p$ as the deviation from the 
ground state ${n}_p^0$.~\cite{rf:PN}\
Hence the physical meaning of ${\varphi}_p$ and $\psi_p$
is now clarified
with the help of eqs.~(\ref{delbarn}) and (\ref{deln}).

It is noted that the total current
$j_\mu(k,\omega)$ given in eq.~(\ref{sigma})
is written also as
\begin{equation}
j_\mu(k,\omega)
=
\frac{e}{\Omega}\sum_{p}
\delta {n}_p j^*_{p\mu},
\label{j=nj}
\end{equation}
with
\begin{equation}
j^*_{p\mu}\equiv
v^*_{p\mu}-
\sum_{p'}
f(\mib{p},\mib{p'})
\frac{\partial n^0_p}{\partial \varepsilon^*_{p'}}
v^*_{p'\mu}.
\end{equation}
Hence $j^*_{p\mu}$  is regarded as the current 
carried by the quasiparticle $p$.
For $j^*_{p\mu}$ thus defined,  
using eq.~(\ref{f=A-Af}),
we find a relation
\begin{eqnarray}
v^*_{p\mu}&\equiv&
j^*_{p\mu}+
\sum_{p'}
A(\mib{p},\mib{p'})
\frac{\partial n^0_p}{\partial \varepsilon^*_{p'}}
j^*_{p'\mu}.
\quad (\omega\gg \gamma^*_p)
\label{v=j+Aj}
\end{eqnarray}
In particular, for $k=0$ and $\omega\gg \gamma^*_p$,
eq.~(\ref{boltz}) without $I({\varphi})$ gives
\begin{eqnarray}
e\mib{v^*_p\cdot E}&=&
-\mbox{i}\omega\psi_p\nonumber\\
&=&-\mbox{i}\omega{\varphi}_p
+
\sum_{p'}
A(\mib{p},\mib{p'})
\frac{\partial n^0}{\partial \varepsilon^*_{p'}}
(-\mbox{i}\omega{\varphi}_{p'}).
\label{inteq}
\end{eqnarray}
Comparing eq.~(\ref{v=j+Aj}) with eq.~(\ref{inteq}),
we can immediately solve the latter for ${\varphi}_p$,
that is, the solution is
\begin{equation}
-\mbox{i}\omega{\varphi}_p
=e\mib{j^*_p\cdot E}.
\label{-iomvp=}
\end{equation}
Therefore, from eqs.~(\ref{sigma}) and (\ref{delbarn}),
we obtain
\begin{equation}
\sigma_{\mu\nu}(\omega)=
\frac{\mbox{i}e^2}{\omega}
\frac{1}{\Omega}\sum_{p'}
v^*_{p\mu}j^*_{p\nu}
\left(-\frac{\partial n^0}{\partial \varepsilon^*_{p}}\right).
\quad (\omega\gg \gamma^*_p)
\label{Drude}
\end{equation}
This is the result of our previous paper.~\cite{rf:I}\

\subsection{Imaginary part of $T^{(0)}$}
\subsubsection{Hydrodynamic regime, $\omega\ll \gamma^*_p$}
\begin{figure}[t]
\centerline{\epsfile{file=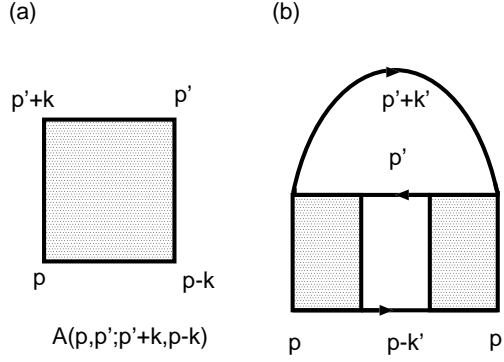,width=6.6cm}}
\caption{(a) The vertex function $A(p,p';p'+k,p-k)$ 
and (b) the selfenergy $\Sigma(p)$ 
which gives rise to the quasiparticle damping $\gamma^*_p\propto T^2$.
}
\label{fig:SE}
\end{figure}
In this subsection we shall 
investigate $I(\varphi)$ of eq.~(\ref{I(varphi)=}).
Among other mechanisms which would contribute to $I({\varphi})$,
we must at least
take account of electron-electron interaction,
which is responsible for making renormalized quasiparticles.
To this end, it is convenient to define
\fulltext
\begin{equation}
\left|A(p,p';p'+k,p-k)\right|^2 
\equiv
z_p z_{p'}z_{p'+k}z_{p-k}
\left|\mbox{Re}\,\Gamma\left(
p-\frac{k}{2},p'+\frac{k}{2};k
\right)\right|^2.
\label{App'p'+kp-k}
\end{equation}
To estimate the term of order $O(T^2)$ for 
the decay rate $\gamma^*_p$,
we have to consider the selfenergy shown in  Fig.~\ref{fig:SE} (b), 
where all the energy variables of $A(p,p';p'+k,p-k)$,
Fig.~\ref{fig:SE} (a), are to be fixed at the Fermi level.
Therefore the function $A(p,p';p'+k,p-k)$ 
which appears in the following equations
is to be regarded as the scattering amplitude of quasiparticle.
Now it is straightforward to obtain~\cite{rf:YY}\
\fulltext
\begin{eqnarray}
2\gamma^*_p
&=&-2z_p\mbox{Im}\,\Sigma^R(\mib{p},0)\nonumber\\
&=&2z_p\int^\infty_{-\infty} \frac{\mbox{d}^4 p'}{(2\pi)^4}
\int^\infty_{-\infty} \frac{\mbox{d}^4  k'}{(2\pi)^4}
\left(\coth\frac{\omega'}{2T}
+\tanh\frac{\omega'}{2T}\right)
\left(\tanh\frac{\varepsilon'}{2T}
-\tanh\frac{\varepsilon'+\omega'}{2T}\right)\nonumber\\
&&\times
\left.
\left|\mbox{Re}\,\Gamma\left(
p-\frac{k}{2},p'+\frac{k}{2};k
\right)
\right|^2 
\mbox{Im}\,G^R(p-k')\mbox{Im}\,G^R(p')\mbox{Im}\,G^R(p'+k')
\right|_{\varepsilon=0}
\nonumber\\
&=&-2\int^\infty_{-\infty} \frac{\mbox{d}^4 p'}{(2\pi)^4}
\int^\infty_{-\infty} \frac{\mbox{d}^4  k'}{(2\pi)^4}
\left(\coth\frac{\omega'}{2T}
-\tanh\frac{\omega'}{2T}\right)
\left(\tanh\frac{\varepsilon'+\omega'}{2T}
-\tanh\frac{\varepsilon'}{2T}\right)\nonumber\\
&&\times\left|A(p,p';p'+k',p-k')\right|^2 
\pi^3
\rho^*_{p-k'}(-\omega')
\rho^*_{p'}(\varepsilon')
\rho^*_{p'+k'}(\varepsilon'+\omega')\nonumber\\
&=&
(\pi T)^2
\sum_{p',k'}\pi\left|A(p,p';p'+k',p-k')\right|^2 
\rho^*_{p-k'}(0)\rho^*_{p'}(0)\rho^*_{p'+k'}(0),
\label{2gamma}
\end{eqnarray}
where 
we used eq.~(\ref{ImGR=}) and  defined
$\rho^*_p(\varepsilon)$ by
\begin{eqnarray}
\rho^*_p(\varepsilon)&\equiv&\delta(\varepsilon-
\varepsilon^*_p+\mu).
\end{eqnarray}
\begin{figure}[t]
\centerline{\epsfile{file=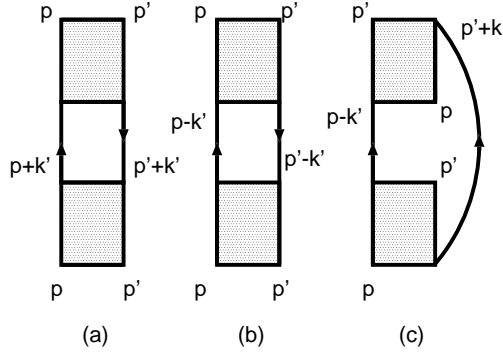,width=6.6cm}}
\caption{
The three irreducible vertices which give
rise to the $T^2$-terms of Im$T^{(0)}_{22}(p,p')$.
These are obtained by cutting off one of the three
solid lines in the selfenergy, Fig.~\ref{fig:SE} (b).
}
\label{fig:ImT0}
\end{figure}
As for Im$T^{(0)}(p,p';k)$, we may set $\mib{k}=0$
to the accuracy of our concern.
To be consistent with the diagram for $\gamma^*_p$,
we have to consider 
three diagrams shown in Fig.~\ref{fig:ImT0}(a) $\sim$ (c).
Considering discontinuity in Im$T^{(0)}(p,p')$ which may appear
at the boundaries between the three regions defined 
in eq.~(\ref{II&III&IV}), we find
\fulltext
\begin{eqnarray}
-\mbox{Im}\Gamma^{{\rm II}}_{a}(p,p')
&=&\mbox{Im}\Gamma^{{\rm III}}_{a}(p,p')
=\mbox{Im}\Gamma^{{\rm IV}}_{a}(p,p')
=\mbox{Im}\,\Gamma^{(a)}(p,p'),\nonumber\\
-\mbox{Im}\Gamma^{{\rm II}}_{b}(p,p')
&=&\mbox{Im}\Gamma^{{\rm III}}_{b}(p,p')
=\mbox{Im}\Gamma^{{\rm IV}}_{b}(p,p')
=\mbox{Im}\,\Gamma^{(b)}(p,p'),\nonumber\\
\mbox{Im}\Gamma^{{\rm II}}_{c}(p,p')
&=&\mbox{Im}\Gamma^{{\rm III}}_{c}(p,p')
=-\mbox{Im}\Gamma^{{\rm IV}}_{c}(p,p')
=\mbox{Im}\,\Gamma^{(c)}(p,p').
\end{eqnarray}
These functions $\Gamma^{(i)}(p,p')$ 
are related to $T^{(i)}(p,p')$
as $\Gamma(p,p')$ is to $T(p,p')$.
Therefore, eq.~(\ref{T22=mendo}) gives
\fulltext
\begin{equation}
\mbox{Im}\,T^{(i)}(p,p')
=-2\left(\coth\frac{\varepsilon'-\varepsilon}{2T}
-\tanh\frac{\varepsilon'}{2T}\right)
\mbox{Im}\,\Gamma^{(i)}(p,p'),
\qquad (i=a,b)
\label{ImTab}
\end{equation}
and
\begin{equation}
\mbox{Im}\,T^{(c)}(p,p')
=2\left(
\coth\frac{\varepsilon'+\varepsilon}{2T}
-\tanh\frac{\varepsilon'}{2T}\right)
\mbox{Im}\,\Gamma^{(c)}(p,p').
\label{ImTc}
\end{equation}
Hence, using
\begin{eqnarray}
\mbox{Im}\,\Gamma^{(a)}(p,p')
&=&
\int \frac{\mbox{d}^4 k'}{(2\pi)^4}
\left(
\tanh\frac{\varepsilon+\omega'}{2T}
-\tanh\frac{\varepsilon'+\omega'}{2T}\right)\nonumber\\
&&\times\left
|\mbox{Re}\,\Gamma\left(p+\frac{k'}{2},p'+\frac{k'}{2};k'
\right)\right|^2 
\mbox{Im}\,G^R(p+k')
\mbox{Im}\,G^R(p'+k')
\end{eqnarray}
we obtain
\begin{eqnarray}
\frac{1}{2}
&\int&\frac{\mbox{d}^4 p'}{(2\pi)^3}
\delta(\varepsilon'-\varepsilon^*_{p'}+\mu)
z_p z_{p'}
\left.
\mbox{Im}\,T^{(a)}(p,p')\right|_{\varepsilon=0}{\varphi}_{p'}
\nonumber\\
&=&2
\int^\infty_{-\infty} \frac{\mbox{d} \varepsilon'}{2\pi}
\int^\infty_{-\infty} \frac{\mbox{d} \omega'}{2\pi}
\left(\coth\frac{\varepsilon'}{2T}
-\tanh\frac{\varepsilon'}{2T}\right)
\left(
\tanh\frac{\varepsilon'+\omega'}{2T}
-\tanh\frac{\omega'}{2T}
\right)\nonumber\\
&&\times
\sum_{p',k'}
\left|A(p,p'+k';p+k',p')\right|^2 
\pi^3
\rho^*_{p'}(\varepsilon')
\rho^*_{p+k'}(\omega')
\rho^*_{p'+k'}(\varepsilon'+\omega')
{\varphi}_{p'}\nonumber\\
&=&
{(\pi T)^2}
\sum_{p',k'}
\pi\left|A(p,p'+k';p+k',p')\right|^2 
\rho^*_{p'}(0)\rho^*_{p+k'}(0)\rho^*_{p'+k'}(0)
{\varphi}_{p'}\nonumber\\
&=&
(\pi T)^2
\sum_{p',k'}
\pi\left|A(p,p';p'+k',p-k')\right|^2 
\rho^*_{p-k'}(0)\rho^*_{p'+k'}(0)\rho^*_{p'}(0)
{\varphi}_{p-k'}.
\label{(a)}
\end{eqnarray}
Similarly, we get
\begin{eqnarray}
\frac{1}{2}
&\int&\frac{\mbox{d}^4 p'}{(2\pi)^3}
\delta(\varepsilon'-\varepsilon^*_{p'}+\mu)
z_p z_{p'}
\left.
\mbox{Im}\,T^{(b)}(p,p')\right|_{\varepsilon=0}{\varphi}_{p'}
\nonumber\\
&=&
{(\pi T)^2}
\sum_{p',k'}
\pi\left|A(p,p'-k';p',p-k')\right|^2 
\rho^*_{p'}(0)\rho^*_{p-k'}(0)\rho^*_{p'-k'}(0)
{\varphi}_{p'}\nonumber\\
&=&{(\pi T)^2}
\sum_{p',k'}
\pi\left|A(p,p';p'+k',p-k')\right|^2 
\rho^*_{p'+k'}(0)\rho^*_{p-k'}(0)\rho^*_{p'}(0)
{\varphi}_{p'+k'},
\label{(b)}
\end{eqnarray}
and
\begin{eqnarray}
\mbox{Im}\,\Gamma^{(c)}(p,p')
&=&-
\int^\infty_{-\infty} \frac{\mbox{d} \omega'}{2\pi}
\int \frac{\mbox{d}^3 k'}{(2\pi)^3}
\left(
\tanh\frac{\varepsilon-\omega'}{2T}
+\tanh\frac{\varepsilon'+\omega'}{2T}\right)\nonumber\\
&&\times\left
|\mbox{Re}\,\Gamma\left(
p-\frac{k'}{2},p'+\frac{k'}{2}; k'
\right)\right|^2 
\mbox{Im}\,G^R(p-k')
\mbox{Im}\,G^R(p'+k'),
\end{eqnarray}
\begin{eqnarray}
\frac{1}{2}
&\int&\frac{\mbox{d}^4 p'}{(2\pi)^3}
\delta(\varepsilon'-\varepsilon^*_{p'}+\mu)
z_p z_{p'}
\left.
\mbox{Im}\,T^{(c)}(p,p')\right|_{\varepsilon=0}{\varphi}_{p'}
\nonumber\\
&=&-2
\int^\infty_{-\infty} \frac{\mbox{d} \varepsilon'}{2\pi}
\int^\infty_{-\infty} \frac{\mbox{d} \omega'}{2\pi}
\left(\coth\frac{\varepsilon'}{2T}
-\tanh\frac{\varepsilon'}{2T}\right)
\left(
\tanh\frac{\varepsilon'+\omega'}{2T}
-\tanh\frac{\omega'}{2T}
\right)\nonumber\\
&&\times
\sum_{p',k'}
\left|A(p,p';p'+k',p-k')\right|^2 
\pi^3
\rho^*_{p'}(\varepsilon')
\rho^*_{p-k'}(-\omega')
\rho^*_{p'+k'}(\varepsilon'+\omega')
{\varphi}_{p'}\nonumber\\
&=&-{(\pi T)^2}
\sum_{p',k'}
\pi\left|A(p,p';p'+k',p-k')\right|^2 
\rho^*_{p'}(0)\rho^*_{p-k'}(0)\rho^*_{p'+k'}(0)
{\varphi}_{p'}.
\label{(c)}
\end{eqnarray}
To take account of quasiparticle spin,
we just have to use
$\left|A_{\sigma\sigma'}(p,p';p'+k',p-k')\right|^2$,
which depends on the spin components $\sigma$ and $\sigma'$
of the incoming quasiparticles $p$ and $p'$,
that is to say, $\left|A\right|^2 $ is to be replaced by 
\halftext
\[
|A_{\uparrow\downarrow}|^2+\frac{1}{2}|A_{\uparrow\uparrow}|^2,\]
where the factor 1/2 in front of $|A_{\uparrow\uparrow}|^2$ 
is added not to overcount the identical quasiparticle states,
namely, 
($p_1\uparrow,p_2\uparrow$) and ($p_2\uparrow,p_1\uparrow$).
As a result, 
with eqs.~(\ref{2gamma}), (\ref{(a)}), (\ref{(b)}) and
(\ref{(c)}) substituted into eq.~(\ref{I(varphi)=}),
we find
\fulltext
\begin{eqnarray}
I({\varphi})&=&-{(\pi T)^2}
\sum_{p',k'}
W(p,p';p'+k',p-k')
\rho^*_{p'}\rho^*_{p-k'}\rho^*_{p'+k'}
({\varphi}_{p}+{\varphi}_{p'}
-{\varphi}_{p'+k'}-{\varphi}_{p-k'}),
\label{I(varphi)}
\end{eqnarray}
\halftext
where $\rho^*_{p}\equiv \rho^*_{p}(0)$.
We defined the transition probability
for the binary collision of quasiparticles,
\begin{equation}
2W\equiv
2\pi\left(
\left|A_{\uparrow\downarrow}\right|^2 
+\frac{1}{2}
\left|A_{\uparrow\uparrow}\right|^2 
\right).
\label{W}
\end{equation}
In effect, eq.~(\ref{I(varphi)}) is obtained from 
the classical formula,
\fulltext
\begin{eqnarray}
I&=&-
\sum_{p',k'}
W(p,p';p'+k',p-k')
\delta(\bar{\varepsilon}^*_p+\bar{\varepsilon}^*_{p'}
-\bar{\varepsilon}^*_{p'+k'}-\bar{\varepsilon}^*_{p-k'})\nonumber\\
&&\times\left[
 \bar{n}_p \bar{n}_{p'}(1-\bar{n}_{p'+k'})(1-\bar{n}_{p-k'})
-(1-\bar{n}_p)(1- \bar{n}_{p'})\bar{n}_{p'+k'}\bar{n}_{p-k'}\right],
\label{Iclassical}
\end{eqnarray}
\halftext
after linearization with respect to 
the deviation $\delta \bar{n}_p$ from the local equilibrium;
\[I\longrightarrow 
\frac{\partial n^0_p}{\partial \varepsilon^*_p}
I({\varphi}).
\]

To summarize, on the basis of the linear response theory
and the finite temperature formalism,
we derived eq.~(\ref{I(varphi)}), 
the collision integral of
the linearized Boltzmann equation (\ref{boltz}).~\cite{comment4}\

\halftext
\subsubsection{Incoherent absorption}
As discussed in the previous paper,~\cite{rf:I}\
the part in the Drude weight that is lost by Umklapp processes
is transfered to the incoherent part $\sigma_{\rm inc}(\omega)$.
Although the full $\omega$-dependence of 
$\sigma_{\rm inc}(\omega)$ cannot be generally described 
in the framework of the Fermi liquid theory,
the low frequency limit of the part due to quasiparticle scattering
can be derived as follows. 
To this end, we consider the correction of order $O(\omega^2)$ 
which should be added to eq.~(\ref{boltz}).

The vertex functions $T_{ij}(p,p';k)$ of eq.~(\ref{Ti2})
(i.e., for $i$ or $j=1,3$)  can be neglected 
since they should appear pairwise and thus if included
give rise to terms of order $O(\omega^3)$ at most.
In eq.~(\ref{g2=}) for $g_2(p,k)$,
we have to consider the energy dependence of $\gamma_p^*$;
\[
\gamma_p^*(\varepsilon)\equiv -z_p\mbox{Im}\,
\Sigma^R(\mib{p},\varepsilon).
\]
Then, $\varepsilon$
in the delta function of  eq.~(\ref{g2=})
should be replaced with $\varepsilon-\omega/2$.
Accordingly, 
we must use $\gamma_p^*(\omega/2)$ and 
$\left.\mbox{Im}\,T^{(0)}(p,p';k)
\right|_{\varepsilon'=\varepsilon=\omega/2}$
in place of 
$\gamma_p^*$  and
$\left.\mbox{Im}\,T^{(0)}(p,p';k)\right|_{\varepsilon=0}$
in eq.~(\ref{I(varphi)=}).
On the other side, for $\mbox{Im}\,T^{(0)}$,
the $\omega$-dependence 
occurs only in the hyperbolic functions
of eqs.~(\ref{ImTab}) and (\ref{ImTc}),
because of the assumption
$\mbox{Im}\,T^{(0)}(p,p';k)=0$ for $\omega=T=0$.
Thus as a result we find~\cite{rf:YY}
\fulltext
\begin{equation}
I({\varphi})=-\left((\omega/2)^2+(\pi T)^2\right)
\sum_{p',k'}
W_{pp'k}
\rho^*_{p'}\rho^*_{p-k'}\rho^*_{p'+k'}
({\varphi}_{p}+{\varphi}_{p'}
-{\varphi}_{p'+k'}-{\varphi}_{p-k'}).
\label{I(varphi)2}
\end{equation}
\halftext
where $W_{pp'k}\equiv W(p,p';p'+k',p-k')$.
It is noted that this is obtained 
from $I$ of eq.~(\ref{Iclassical})
by formally using 
$\delta(\bar{\varepsilon}^*_p+\bar{\varepsilon}^*_{p'}
-\bar{\varepsilon}^*_{p'+k'}-\bar{\varepsilon}^*_{p-k'}+\omega)$
instead of the delta function in eq.~(\ref{Iclassical}).
The collision integral $I(\varphi)$ of eq.~(\ref{I(varphi)2})
is used to describe damping of zero sound.~\cite{rf:Landau}\
Therefore, our microscopic derivation proves that
the quasiparticle part of 
the matrix element $W(p,p';p'+k',p-k')$
of the zero sound damping  coincides with 
the transition probability 
for the binary collision of quasiparticles.

To discuss the quasiparticle part of the optical conductivity,  
we assume $k=0$ and $T=0$.
Then eq.~(\ref{boltz}) reads
\fulltext
\begin{eqnarray}
-\mbox{i}\omega\psi_p+
e\mib{v^*_p\cdot E}&=&-\frac{\omega^2}{4}\sum_{p',k'}
W_{pp'k'}
\rho^*_{p'}\rho^*_{p-k'}\rho^*_{p'+k'}
({\varphi}_{p}+{\varphi}_{p'}
-{\varphi}_{p'+k'}-{\varphi}_{p-k'}).
\label{Case;k0T0}
\end{eqnarray}
\halftext
This is solved by successive approximation
as a series in powers of $\omega$:
\begin{eqnarray}
\mbox{i}\omega\psi_p&=&
\mbox{i}\omega\psi_p^{(0)}
+\mbox{i}\omega\psi_p^{(1)},\nonumber\\
\mbox{i}\omega\varphi_p&=&
\mbox{i}\omega\varphi_p^{(0)}
+\mbox{i}\omega\varphi_p^{(1)},
\end{eqnarray}
where 
\[\mbox{i}\omega\psi_p^{(0)}\sim O(1),\quad
\mbox{i}\omega\psi_p^{(1)}\sim O(\omega),
\quad{\rm etc.}
\]
The terms in this series satisfy the equations,
\begin{equation}
-\mbox{i}\omega\psi_p^{(0)}+e\mib{v^*_p\cdot E}=0,
\end{equation}
and
\fulltext
\begin{equation}
-\mbox{i}\omega\psi_p^{(1)}=
-\frac{\omega^2}{4}\sum_{p',k'}
W_{pp'k'}
\rho^*_{p'}\rho^*_{p-k'}\rho^*_{p'+k'}
({\varphi}_{p}^{(0)}+{\varphi}_{p'}^{(0)}
-{\varphi}_{p'+k'}^{(0)}-{\varphi}_{p-k'}^{(0)}).
\end{equation}\halftext
As in eq.~(\ref{-iomvp=}), the former gives
\begin{equation}
-\mbox{i}\omega{\varphi}_p^{(0)}=e\mib{j^*_p\cdot E}.
\end{equation}
Then, the latter gives 
\fulltext
\begin{equation}
-\mbox{i}\omega\psi_p^{(1)}=
\frac{\mbox{i}e\omega}{4}\sum_{p',k'}\sum_\mu
W_{pp'k'}
\rho^*_{p'}\rho^*_{p-k'}\rho^*_{p'+k'}
({j^*_{p\mu}} +{j^*_{p'\mu}}
-{j^*_{p'+k'\mu}}-{j^*_{p-k'\mu}})E_{\mu}.
\end{equation}
\halftext
Therefore, we obtain
\begin{equation}
\sigma_{\mu\nu}(\omega)
=\sigma^{(0)}_{\mu\nu}+\sigma^{(1)}_{\mu\nu},
\end{equation}
for which 
$\sigma_{\mu\nu}^{(0)}$ is given by eq.~(\ref{Drude})
and
$\sigma_{\mu\nu}^{(1)}$ is 
\begin{equation}
\sigma_{\mu\nu}^{(1)}=
\frac{e^2}{4\Omega}\sum_{p,p',k'}
W_{pp'k'}
\rho^*_{p}\rho^*_{p'}\rho^*_{p-k'}\rho^*_{p'+k'}
j^*_{p\mu}({j^*_{p\nu}} +{j^*_{p'\nu}}
-{j^*_{p'+k'\nu}}-{j^*_{p-k'\nu}}),
\label{sig1}
\end{equation}
\halftext
where we used eqs.~(\ref{deln}) and (\ref{j=nj})
for $\psi_p^{(1)}$, and
\[
-\frac{\partial n^0_p}{\partial \varepsilon^*_p}
=\rho^*_p.
\]
Equation~(\ref{sig1}) indicates 
the incoherent part $\sigma^{(1)}_{\mu\nu}$ must vanish 
unless the current conservation is violated.
Thus we obtain $\sigma^{(1)}_{\mu\nu}=0$  
in an isotropic system, as it should be the case.

\section{$\mib{T^2}$ Contribution to the Electrical
Resistivity}
\subsection{Formula}

In this section, we discuss 
the dc conductivity 
due to quasiparticle transport at finite temperature.
For $\omega=0$ and $k=0$, eq.~(\ref{boltz}) 
with eq.~(\ref{I(varphi)}) gives
\fulltext
\begin{eqnarray}
e\mib{v^*_p\cdot E}&=&-{(\pi T)^2}
\sum_{p',k'}
W_{pp'k'}
\rho^*_{p'}\rho^*_{p-k'}\rho^*_{p'+k'}
({\varphi}_{p}+{\varphi}_{p'}
-{\varphi}_{p'+k'}-{\varphi}_{p-k'}).
\label{Case;om0k0}
\end{eqnarray}
Yamada and Yosida~\cite{rf:YY} were the first who 
derived this sort of equation
on the basis of the linear response theory.
In their work, the function 
${\varphi}_{p}$ was defined after 
they derived the integral equation for 
the full vertex part $\tilde{\mib{{\Lambda}}}_p$
that includes the effect of
the section $g_2$ as well as $g_1$ and $g_3$;
in our notation,  they defined ${\varphi}_{p}$ by
\halftext
\begin{equation}
{\varphi}_{p}\equiv
-\frac{e\tilde{\mib{\Lambda}}_p\mib{\cdot E}}{2\Delta_p},
\label{varphiYY}
\end{equation}
where
\[\Delta_p=-\mbox{Im}\,\Sigma^R(\mib{p},0).\]
Eq.~(\ref{varphiYY}) is to be compared with 
eq.~(\ref{varphi0}) by noting
\begin{equation}
g_2(p,0)
=\frac{\pi z_p}{\Delta_p}
\delta(\varepsilon-\varepsilon^*_p+\mu).
\end{equation}
As discussed above, our derivation 
gives a definite interpretation of ${\varphi}_{p}$
as the deviation from the local equilibrium distribution function,
eq.~(\ref{delbarn}).

\begin{figure}[t]
\centerline{\epsfile{file=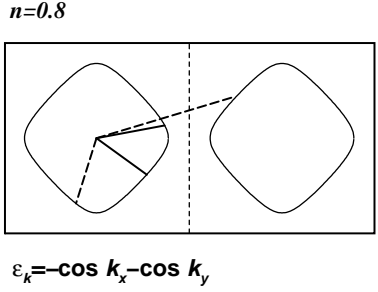,width=6.6cm}}
\caption{
An example of Umklapp processes on the Fermi surface of
$\varepsilon_k=-\cos k_x-\cos k_y$ for $n=0.8$.
The initial vectors $k_1$ and $k_2$ (solid lines)
are scattered to $k_3$ and $k_4$ (dashed lines).
}
\label{fig:umklapp}
\end{figure}

As is usually the case in solving the Boltzmann equation,
we must assume a specific form for the function ${\varphi}_{p}$
to solve eq~(\ref{Case;om0k0}).
In an isotropic system,
as a scalar function of the vectors $\mib p$ and $\mib E$,  
the function $\varphi_p$ is assumed to be
$\varphi_p\propto \mib{p\cdot E}$.
Then the right-hand side of 
eq.~(\ref{Case;om0k0})  identically vanishes
owing to momentum conservation,
indicating that the current persists
even at finite temperatures.~\cite{rf:YY}\
In general, on physical grounds, we may assume 
 \[{\varphi}_{p}=-e\tau^*_p\mib{v^*_p\cdot E},\]
with $\tau^*_p$ as a function of $p$.
However, for simplicity, 
neglecting the momentum dependence of $\tau^*_{p}$,
we assume
\begin{equation}
{\varphi}_{p}=-e\tau^*_{\rm tr}\mib{v^*_p\cdot E}.
\label{varphitau*}
\end{equation}
Then, the transport relaxation time $\tau^*_{\rm tr}$
is obtained from eq.~(\ref{Case;om0k0}); 
\fulltext
\begin{eqnarray}
\frac{1}{\tau^*_{\rm tr}}&=&
(\pi T)^2
\frac{
\displaystyle 
\sum_{p,p',k}
W_{pp'k}
\rho^*_{p}\rho^*_{p'}\rho^*_{p-k}\rho^*_{p'+k}
 v^*_{p\nu}
(v^*_{p\nu}+v^*_{p'\nu}
-v^*_{p'+k\nu}-v^*_{p-k\nu})}
{\displaystyle 
\sum_p \rho^*_{p}v^*_{p\nu}v^*_{p\nu}}.
\label{1/tautr}
\end{eqnarray} 
\halftext
It is stressed here that this result based on eq.~(\ref{varphitau*})
is approximate.
One may follow a recent work of
Maebashi and Fukuyama~\cite{rf:MF1}
to improve upon the approximation systematically.
It is interesting to see that
the velocity $v^*_{p}$ in eq.~(\ref{1/tautr}) take the place
of the current $j^*_{p}$ of eq.~(\ref{sig1})
in the coherent regime.
By eqs.~(\ref{sigma}) and (\ref{delbarn}),
the conductivity $\sigma_{\nu\nu}(0)$ is written
\begin{eqnarray}
\sigma_{\nu\nu}(0)=
\frac{e^2}{\Omega}
\sum_{p\sigma}
\rho^*_{p}v^*_{p\nu}v^*_{p\nu}\tau^*_{\rm tr}.
\end{eqnarray}
Hence, the coefficient $A$
of the electrical resistivity $R=\sigma_{\nu\nu}(0)^{-1}=AT^2$
is given by
\fulltext
\begin{equation}
\frac{2e^2A}{\pi^2}=
\frac{\displaystyle
\frac{1}{\Omega}
\sum_{p,p',k}
W_{pp'k}
\rho^*_{p}\rho^*_{p'}\rho^*_{p-k}\rho^*_{p'+k}
 v^*_{p\nu}
(v^*_{p\nu}+v^*_{p'\nu}
-v^*_{p'+k\nu}-v^*_{p-k\nu})
}{\displaystyle
\left(\frac{1}{\Omega}
\sum_p \rho^*_{p}v^*_{p\nu}v^*_{p\nu}\right)^2}.
\label{A}
\end{equation}
\halftext
In our simple cases discussed below,
the normal processes which give finite contribution 
to the summation,
\begin{equation}
\sum_{p_1,p_2,p_3}\rho_{p_1}\rho_{p_2}\rho_{p_3}\rho_{p_4}
\delta_{p_1+p_2-p_3-p_4},
\end{equation}
are those satisfying (i) $p_1=p_3$, $p_2=p_4$,
(ii) $p_1=p_4$, $p_2=p_3$ or (iii) $p_1+p_2=p_3+p_4=0$.
Therefore, the normal processes do not contribute 
to $A$ because of the factor 
$v^*_{p\nu}+v^*_{p'\nu}-v^*_{p'+k\nu}-v^*_{p-k\nu}=0$
in eq.~(\ref{A}):
Non-zero resistivity is brought about by 
Umklapp processes~\cite{comment}\ (Fig.~\ref{fig:umklapp}).
Moreover, eq.~(\ref{A}) shows
that we must assume
$W_{pp'k}\propto |A_{pp'k}|^2 \propto (\rho^*)^{-2}$
so as to obtain
the Kadowaki-Woods relation $A\propto (v^*)^{-2}\propto (\rho^*)^2$.
In this respect,
we can make an order of magnitude estimate of $\rho^* A_{pp'k}$
in terms of eq.~(\ref{f=A-Af}),
which shows that
$\rho^* A(p,p')$ in fact
approaches a constant independent of $\rho^* $ 
in the strong coupling regime $\rho^* \Omega f(p,p')\gg 1$:
It is instructive to refer eq.~(\ref{A_l^s=F}) by which
we obtain $A_l\simeq \rho^* \Omega A(p,p')\ \alt O(1)\ $ for $F_l\gg 1$.
Thus in this situation
we are led to the Kadowaki-Woods relation $A\propto \gamma^2$
with $\gamma$ of the specific heat coefficient,
\begin{equation}
\gamma=\frac{2\pi^2}{3}\rho^*,\qquad
\rho^*=
\frac{1}{\Omega}
\sum_{p}\rho_p^*.
\label{gamma}
\end{equation}
The ratio $A/\gamma^2$ 
however may depend on the geometry of the Fermi surface
relative to the Brillouin zone boundary.
To give a rough estimate, 
neglecting the momentum dependence of $W_{pp'k}$ or 
$A_{pp'k}$,
we introduce the coupling constants $A^s$ and $A^a$ by
\begin{eqnarray}
2 \rho^* \Omega
A_{\uparrow\downarrow}\equiv A^s-A^a, &\quad&
2 \rho^* \Omega A_{\uparrow\uparrow}\equiv A^s+A^a.
\label{2rho*A}
\end{eqnarray}
Then, we obtain
\begin{eqnarray}
\frac{A}{\gamma^2}
&=&\frac{9\hbar a}{8e^2\pi} \alpha F\nonumber\\
&=&8.5 \alpha F
\times 10^{-7} \mu {\Omega}\mbox{cm}(\mbox{mol}\,\mbox{K/mJ})^2,
\qquad (a={\rm 4\AA})
\label{A/gamma2}
\end{eqnarray}
where
\begin{equation}
\alpha\equiv
\frac{1}{4}\left( (A^s-A^a)^2+\frac{1}{2}(A^s+A^a)^2\right)
\equiv\alpha_{\uparrow\downarrow}+\alpha_{\uparrow\uparrow},
\label{alpha}
\end{equation}
and
\fulltext
\begin{equation}
F=\frac{\displaystyle
\frac{1}{\Omega^3}\sum_{p,p',k}
\rho^*_{p}\rho^*_{p'}\rho^*_{p-k}\rho^*_{p'+k}
 v^*_{p\nu}
(v^*_{p\nu}+v^*_{p'\nu}
-v^*_{p'+k\nu}-v^*_{p-k\nu})
}{\displaystyle (\rho^*)^4
\left(\frac{1}{\Omega}\sum_p \rho^*_{p}v^*_{p\nu}v^*_{p\nu}\right)^2}.
\label{F}
\end{equation}
\halftext
The parameter $\alpha$ of eq.~(\ref{alpha})
is a coupling constant of quasiparticle interaction.
On the other side,
the parameter $F$ is invariant under the transformation,
\begin{eqnarray} 
\varepsilon^*_p-\mu &\rightarrow& z (\varepsilon^*_p-\mu), \nonumber\\
v^*_p &\rightarrow& z v^*_p,\nonumber\\
\rho^*_p&\rightarrow &\rho^*_p/z.
\end{eqnarray}
Thus $F$ is regarded as a quantity characterizing 
a shape of the Fermi surface,
so that we may disregard
the superscript * for the renormalized quantities in eq.~(\ref{F}).
Equation~(\ref{A/gamma2}) indicates that
for the systems obeying the Kadowaki-Woods relation
the product $\alpha F$  must be a universal constant.

\subsection{Two-dimensional systems}
To estimate $F$ quantitatively,
it is convenient to introduce a coordinate system on the Fermi surface.
Although it is possible
to treat an arbitrary case in principle,
for simplicity,
we consider two dimensional systems with a closed Fermi `surface'
in the Brillouin zone ($-\pi\le k_x, k_y\le \pi$).
Since then points on the Fermi surface are parameterized by 
the azimuthal angle $\theta=\tan^{-1}k_y/k_x $, 
we define the function  $k_\theta$, which is determined 
so as to satisfy \(\varepsilon_k=\mu\),
for $(k_x,k_y)=(k_\theta\cos\theta,k_\theta\sin\theta)$.
In terms of the function $k_\theta$,
local density of states on the Fermi surface is given by
\begin{equation}
\rho_\theta\mbox{d}\theta=
\frac{1}{(2\pi)^2}\frac{\sqrt{k_\theta^2+\dot{k}_\theta^2}}
{v_\theta}\mbox{d}\theta,
\end{equation}
where
\[\dot{k}_\theta=\frac{\mbox{d} k_\theta}{\mbox{d} \theta},
\quad 
v_\theta=\sqrt{v_x^2+v_y^2},
\quad
v_\mu=\frac{\partial \varepsilon_k}{\partial k_\mu}.
\]
Then, we may use
\[
\frac{1}{\Omega}
\sum_{k}=\int \rho_\theta\mbox{d}\theta\mbox{d}\varepsilon,
\]
in which 
the integral over $\varepsilon$ eliminates the delta function 
\(
\rho_p(\varepsilon)=\delta(\varepsilon-
\varepsilon_p+\mu).\)
Thus we can use
\fulltext
\begin{equation}
\frac{1}{\Omega^3}
\sum_{k_1,k_2,k_3}\rho_{k_1}\rho_{k_2}\rho_{k_3}\rho_{k_4}
\delta_{k_1+k_2-k_3-k_4}
=\int \mbox{d} \theta_1\mbox{d} \theta_2\mbox{d} \theta_3
\rho_{\theta_1}\rho_{\theta_2}\rho_{\theta_3} \rho_{k_4},
\label{sum3}
\end{equation}\halftext
where $k_4=k_1+k_2-k_3$.
We assumed that $k_1$, $k_2$ and $k_3$ are 
on the Fermi surface in the first Brillouin zone,
without loss of generality. 
The process in which to have $k_4$ outside the Brillouin zone is
the Umklapp process
giving rise to a finite contribution to $F$.
To eliminate the last delta function $\rho_{k_4}$
in eq.~(\ref{sum3}),
we shall regard $\varepsilon_{k_4}$
as a function of $\theta_3$,
i.e., $\varepsilon_{k_4}\equiv \epsilon(\theta_3)$,
and define $\bar{\theta}_3$ as a solution of 
$\epsilon(\bar{\theta}_3)=\mu$, 
by which $\bar{\theta}_3$ is 
determined as a function of $\theta_1$ and $\theta_2$.
Then, we may put 
\begin{equation}
\rho_{k_4}=\delta(\theta_3-\bar{\theta}_3)/a_{\bar{\theta}_3}, \quad
a_{\bar{\theta}_3}=\left|\frac{\mbox{d}\epsilon(\bar{\theta}_3)}
{\mbox{d} \theta_3}\right|,\nonumber
\end{equation}
and the delta function
is eliminated by the integral over $\theta_3$.
As a result, to evaluate eq.~(\ref{F}) 
we may use
\fulltext
\begin{equation}
\frac{1}{\Omega^3}
\sum_{k_1,k_2,k_3}\rho_{k_1}\rho_{k_2}\rho_{k_3}\rho_{k_4}
\delta_{k_1+k_2-k_3-k_4}
=
\int\mbox{d} \theta_1\mbox{d} \theta_2
\rho_{\theta_1}\rho_{\theta_2}\rho_{\bar{\theta}_3}
/a_{\bar{\theta}_3}.
\end{equation}
\halftext

\begin{figure}[t]
\centerline{\epsfile{file=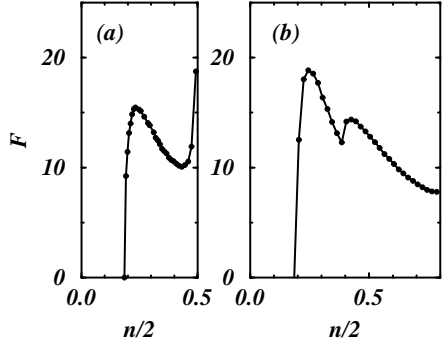,width=6.6cm}}
\caption{$F$ as a function of $n$ for
(a) $\varepsilon_k=-\cos k_x-\cos k_y$ ($n<1$),
and (b) $\varepsilon_k=k_x^2+k_y^2$
($n< \pi/2$).
The function $F$ identically vanishes for $n/2<n_c/2\simeq 0.2$
owing to the absence of Umklapp scattering.
}
\label{fig:F}
\end{figure}
In this manner, we calculated the parameter $F$ 
as a function of density $n$ for 
(a) $\varepsilon_k=-\cos k_x-\cos k_y$ for $n<1$,
and (b) $\varepsilon_k=k_x^2+k_y^2$ for $n<\pi/2$
(Fig.~\ref{fig:F}.)
The result of the case (a) alone 
may be obtained by using the recent results for the Hubbard model 
by Maebashi and Fukuyama.~\cite{rf:MF}
Comparing the two cases,
we see that $F$ as a function of $n$ does not 
depend so much on the shape of the Fermi surface,
except for a singular behavior 
near half filling $n=1$ of the case (a).~\cite{rf:FKY,rf:MF}\
In effect, from the figure, we observe $F\simeq 10$
as far as Umklapp processes are effective.
For small $n$ ($<n_c$), there can be no Umklapp process available,
so that we obtain $F=0$ identically.
It is elementary to estimate the critical concentration $n_c$.
Using $k_4=k_1+k_2-k_3=3k_1$ for
$\mib{k_1}=\mib{k_2}=-\mib{k_3}\parallel \mib{k_4}$,
on the threshold of the Umklapp process $k_4=2\pi-k_1$, 
we obtain $k_1=\pi/2$.
This corresponds to 
$n_c/2=\pi k_1^2/(2\pi)^2=\pi/16=0.196$ for 
$\varepsilon_k=k^2$.
This is consistent with the results shown in the figure.

\subsection{Three-dimensional systems}
\begin{figure}[t]
\centerline{\epsfile{file=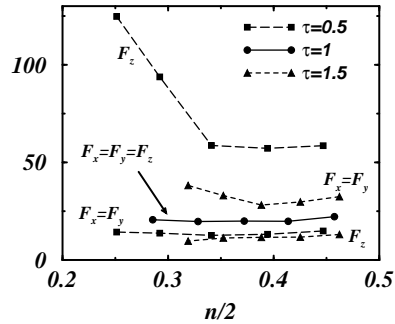,width=6.6cm}}
\caption{
The parameters $F_\mu$ as a function of $n$
for the Fermi surface of
$\varepsilon_k=-\cos k_x-\cos k_y -\tau \cos k_z$.
}
\label{fig:num-F}
\end{figure}
\begin{figure}[t]
\centerline{\epsfile{file=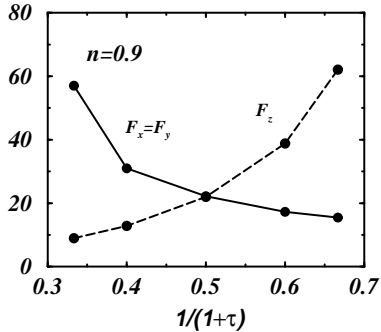,width=6.6cm}}
\caption{
The parameters $F_\mu$ for $n=0.9$ as a function of
the anisotropy $1/(1+\tau)$ of
$\varepsilon_k=-\cos k_x-\cos k_y -\tau \cos k_z$.
}
\label{fig:tau-F}
\end{figure}
Next we consider
the Fermi surfaces of three dimensional systems
which may be parameterized by cylindrical coordinates
$\mib{k}=(k\cos\theta,k\sin\theta, k_z)$,
where the radial function $k$ is now regarded as a 
function of $\theta $ and $k_z$.
Then we can proceed similarly as in the previous subsection,
i.e., we use
\[
\frac{1}{\Omega}
\sum_{k}=\int \rho 
\mbox{d}\theta\mbox{d}k_z \mbox{d}\varepsilon,
\]
where
\begin{equation}
\rho 
\mbox{d}\theta\mbox{d}k_z=
\frac{1}{(2\pi)^3}
\sqrt{k^2 (1+\partial_z k^2)
+\partial_\theta k^2}\
\frac{\mbox{d}\theta\mbox{d}k_z}{v},
\end{equation}
and
\[
v=\sqrt{v_x^2+v_y^2+v_z^2},
\quad
v_\mu=\frac{\partial \varepsilon_k}{\partial k_\mu}.
\]
We evaluated $F_\mu$ for the dispersion
\[
\varepsilon_k=-\cos k_x-\cos k_y-\tau\cos k_z,
\]
as a function of the density $n$ (Fig.~\ref{fig:num-F})
and the anisotropy $\tau$ (Fig.~\ref{fig:tau-F}).
From the figures we see $F_\mu \alt 20$,
except for those in resistive directions
such as $F_z$ for $\tau<1$
and $F_x=F_y$ for $\tau>1$.

To conclude this section, 
let us estimate the `universal' ratio $A/\gamma^2$.
As noted above, the two parameters 
$\alpha$ and $F_\mu$ should take constant values
for the ratio $A_\mu/\gamma^2$ to be universal.
For the former we may assume the strong coupling condition
$\alpha \sim 1$ for the universality.
For the latter, we assume $F=10\sim 20$ as a typical value
indicated by the above figures.
Then, for example, 
for $\alpha=1$ (e.g., for $A^s=1$ and $A^a=-1$) and $F=20$,
eq.~(\ref{A/gamma2}) gives \(A/\gamma^2\simeq 1.7
\times 10^{-5} \mu {\Omega}\mbox{cm}(\mbox{mol}\,\mbox{K/mJ})^2.
\)
This is reasonable as a rough estimate for 
the observed ratio~\cite{rf:KW}
\(1.0
\times 10^{-5} \mu {\Omega}\mbox{cm}(\mbox{mol}\,\mbox{K/mJ})^2
\).

\section{Discussion}

In this article,
on the basis of \'Eliashberg's formalism~\cite{rf:Eliashberg}
and using the results of Yamada and Yosida,~\cite{rf:YY}
we formulated the Landau Fermi liquid theory microscopically
in the form presented by Pines and Nozi\`eres.~\cite{rf:PN} \
In so doing our motivation was to eliminate microscopic quantities
$\Sigma(p)$ and $\Gamma(p,p';k)$
in favor of $\varepsilon^*_k$ and $A_{pp'k}$;
to derive general results
one should not introduce approximations
even if they are found to be physically appropriate.
In a sense we must use $A_{pp'k}$ and $A(\mib{p},\mib{p'})=A_{pp'k=0}$ 
for the theory to be definite or to avoid confusion,
since $\Gamma(\mib{p},\mib{p'})
\equiv\Gamma(\mib{p},\mib{p'};\mib{p'},\mib{p})
\equiv\lim_{k\rightarrow 0}\Gamma(p,p';k)$ 
is not well defined as it depends on the limiting procedure. 
The Landau parameters $f(\mib{p},\mib{p'})$, which are usually used 
to describe static properties of the system,
are obtained from the forward scattering amplitudes 
$A(\mib{p},\mib{p'})$
by solving eq.~(\ref{f=A-Af}).
In effect, for practical purposes,
we do not have to know $f(\mib{p},\mib{p'})$
since physical quantities in the hydrodynamic regime 
are usually 
expressed in terms of $A(\mib{p},\mib{p'})$ in a direct manner.
For example, the total magnetization $M$ 
in the magnetic field $H$ is given by
\begin{equation}
M= \mu_{\rm B}\sum_{p,\sigma=\pm}\sigma \delta n_{p\sigma},
\end{equation}
while the effect of $H$ causes the deviation 
from the local equilibrium,
\begin{equation}
\delta \bar{n}_{p\sigma}=
\mu_{\rm B}\sigma H 
\left(-\frac{\partial n^0}{\partial \varepsilon^*_p}\right).
\end{equation}
Using eq.~(\ref{deln=delbarn+}) to express
$\delta n_{p\sigma}$ in terms of $\delta \bar{n}_{p\sigma}$,
the susceptibility $\chi_{s}$ is given by
\begin{eqnarray}
\chi_{s}&=&\frac{M}{\Omega H}
=2\mu_{\rm B}^2\rho^*(1-A_0^a),
\label{chis}
\end{eqnarray}
where 
\begin{equation}
  \label{2rho*Aa}
2\rho^* 
A_0^a\equiv \frac{4}{\Omega} \sum_{p,p'}
\rho_p^*\rho_{p'}^*A^a(\mib{p},\mib{p'}),
\end{equation}
\begin{equation}
  \label{2Aa}
2A^a(\mib{p},\mib{p'})=
A_{\uparrow\uparrow}(\mib{p},\mib{p'})-
A_{\uparrow\downarrow}(\mib{p},\mib{p'}).
\end{equation}
In particular, for an isotropic system, eq.~(\ref{A_l^s=F}) gives 
\[
A_0^a=\frac{F^a_0}{1+F^a_0},
\]
and we recover a well-known result, 
\[
\chi_{s}=\frac{2\mu_{\rm B}^2\rho^*}{1+F^a_0}.
\]
Similarly, the charge susceptibility $\chi_{c}$ is given by
\begin{equation}
\chi_{c}=\frac{\partial n}{\partial \mu}=2\rho^*(1-A_0^s),
\label{chic}
\end{equation}
\begin{equation}
2\rho^*  A_0^s \equiv \frac{4}{\Omega}
\sum_{p,p'}
\rho_p^*\rho_{p'}^*A^s(\mib{p},\mib{p'}),
\end{equation}
\begin{equation}
2A^s(\mib{p},\mib{p'})=
A_{\uparrow\uparrow}(\mib{p},\mib{p'})+
A_{\uparrow\downarrow}(\mib{p},\mib{p'}).
\end{equation}
It is noted that part of 
Yamada and Yosida's results~\cite{rf:YY} for $\chi_s$ and $\chi_c$ 
due to $f$ electrons in  heavy electron systems 
can be cast into the above forms.
In other words, 
eqs.~(\ref{chis}) and (\ref{chic}) are derived microscopically
by using the results of ref.~\citen{rf:YY}.\
The derivation is as follows.
From eqs.~(2.18) and (2.21) of ref.~\citen{rf:YY},
we get
\begin{equation}
\chi_s=\frac{2\mu_{\rm B}^2}{\Omega}
\sum_p z_p \rho^*_p 
\left(\tilde{\chi}_{\uparrow\uparrow}(\mib{p})
+\tilde{\chi}_{\uparrow\downarrow}(\mib{p})
\right).
\end{equation}
On the other side, 
from eqs.~(3.2), (3.9) and (3.16) of ref.~\citen{rf:YY}, we obtain
\begin{equation}
\rho^*=\frac{1}{\Omega}\sum_p 
z_p \rho^*_p \tilde{\chi}_{\uparrow\uparrow}(\mib{p})
+\frac{1}{\Omega}\sum_{p,p'}
\rho_p^*\rho_{p'}^* A_{\uparrow\uparrow}(\mib{p},\mib{p'}),
\end{equation}
and 
\begin{equation}
\frac{1}{\Omega}\sum_p 
z_p \rho^*_p \tilde{\chi}_{\uparrow\downarrow}(\mib{p})
=\frac{1}{\Omega}\sum_{p,p'}
\rho_p^*\rho_{p'}^* A_{\uparrow\downarrow}(\mib{p},\mib{p'}),
\end{equation}
respectively.
By eliminating 
$\tilde{\chi}_{\uparrow\uparrow}(\mib{p})$
and $\tilde{\chi}_{\uparrow\downarrow}(\mib{p})$,
we are led to eq.~(\ref{chis}) with
eqs.~(\ref{2rho*Aa}) and (\ref{2Aa}).
The result for $\chi_c$ can be derived similarly.
In this formulation, the Wilson ratio $R_{\rm W}$ 
is simply expressed by 
the single parameter $A^0_a$ as $R_{\rm W}=1-A^0_a$,
instead of two parameters 
$\chi_{\uparrow\downarrow}/\chi_{\uparrow\uparrow}$
and $\delta_{\uparrow\uparrow}/\chi_{\uparrow\uparrow}$
as discussed below 
eq.~(3.20) of ref.~\citen{rf:YY}.
This is because of our proper use of $\rho^*$ 
for the specific heat coefficient $\gamma$
without neglecting the momentum dependence of the selfenergy $\Sigma(p)$.

As is clear from these examples and eq.~(\ref{A}), in all cases, 
the scattering amplitude $A_{pp'k}$ enters in physical quantities 
as some kind of average over momentum variables at the Fermi level.
Therefore the whole momentum-dependence of $A_{pp'k}$
would not be determined solely by experiment, or
the theory requires more information than experiment offers.
An ultimate goal on the theoretical side would be to 
derive $\varepsilon^*_k$ and $A_{pp'k}$ 
by eqs.~(\ref{varepsilon*}) and (\ref{App'p'+kp-k})
from first principles, e.g.,
on the basis of the periodic Anderson Hamiltonian.
As this is far from a trivial task,
one has to regard these as parameters in practice,
giving up explaining 
why and to what extent the quasiparticle mass of 
a certain system is heavily enhanced.
Then, to sum up, 
all we can do is to obtain 
an order-of-magnitude estimate of the parameters $A^s$ and $A^a$
from, e.g.,  $A^s_0$ and $A^a_0$ of 
eqs.~(\ref{chis}) and (\ref{chic}).
Nevertheless,  this is no way disappointing in view of the fact 
that there are a large class of systems among which
$\rho^*$ 
may vary 
so widely that the rough estimation can still make sense.
To put it differently,
one may consider that
the observed universality in the ratio $A/\gamma^2$
supports
the assumption to neglect the momentum dependence of 
the scattering amplitude in eq.~(\ref{2rho*A}).

Keeping these points in mind, 
let us discuss factors which may spoil 
the agreement between theory and experiment 
on the ratio $A/\gamma^2$, 
noted at the end of the previous section.
(i) The lattice constant $a$ is assumed to be $a=$4\AA:
The error caused by this factor will not be appreciable.
(ii) On the geometric factor $F$:
We estimated $F$ only for several simple examples.
In our theory,
the assumption that $F \alt 20$ 
will not depend on the detailed structure of the Fermi surface
is crucial to reproduce 
the universal value for the ratio $A/\gamma^2$.
In principle, given the Fermi surface,
the factor $F$ defined in eq.~(\ref{F}) 
can be evaluated
by parameterizing coordinates on the Fermi surface.
Thus it is desirable to make sure that
this is in fact the case for
as many arbitrary but typical Fermi surfaces as possible.
As examples for which $F$ deviates from the universal value,
we showed the results for anisotropic systems in the previous section.
Moreover, we can imagine an exceptional case as 
for $n=1$ of the square lattice~\cite{rf:FKY,rf:MF}
(Fig.~{\ref{fig:F}(a)).
In general, one can define
the geometric factor $F$
for the bands comprising degenerate orbitals. 
In this case,  as a result of possible interband Umklapp processes, 
the geometric factor $F$ may deviate from the universal value
estimated in single band models with large Fermi surfaces.
For example, $F$ can become small 
in the case where intraband Umklapp processes are suppressed
by a peculiar geometry of available phase space,
e.g,  when relevant Fermi surfaces are too small compared with 
the size of unit cell spanned by reciprocal wavevectors.
This point may be relevant to the experimental fact
that the ratio $A/\gamma^2$ for the transition metals
such as Ni and Pd in the vicinity of ferromagnetic instability
is more than an order of magnitude smaller than that for 
the heavy fermion compounds,~\cite{rf:Rice}
although it may be due to a relatively weak electron correlation effect,
or due to small $\alpha$.
In our formulation, the latter possibility may correspond to 
the viewpoint of Miyake $et$ $al.$~\cite{rf:MMV}\
(iii) 
We must have $\alpha\sim 1$:
If we were to have $F\alt 20$, then 
$\alpha\sim 1$ is required to explain the universality
$A/\gamma^2\simeq 
1.0
\times 10^{-5} \mu {\Omega}\mbox{cm}(\mbox{mol}\,\mbox{K/mJ})^2$, 
as observed experimentally.
For this to be the case,
we must be not only in the strong coupling (repulsive) regime 
where $2\rho^* \Omega f(p,p')\gg 1$ (i.e., $A^s\simeq 1$),
but also well away from the Fermi liquid instabilities
so that $|A^a|\sim 1$.
When the quasiparticle interaction is attractive 
and the system is in the close vicinity of 
the Fermi liquid instability,~\cite{rf:comment3}
$|A(p,p')|^2$ can become quite large, 
and the ratio $A/\gamma^2$ can deviate from a universal value.
In this regard, however, a note is in order.
The parameter $A^a_0$ determined from $\chi_s$, for example,
is defined by the forward scattering amplitude  $A^a(p,p')$.
Thus it may happen that
the enhancement which give rise to a large $|A^a_0|$ 
is prominent only in the forward region $k\sim 0$ 
of the scattering amplitude $|A_{pp'k}|$.
Then, the proximity to the instability 
does not necessarily imply a large resistivity 
since those processes with $k\sim 0$ 
do not contribute much to the resistivity.

With respect to the point (iii) noted above,
well-known characteristics  of the heavy fermion systems 
obeying the Kadowaki-Woods relation are regarded 
as general properties of stable strong-coupling Fermi liquids;
that is,
the charge susceptibility $\chi_c$, eq.~(\ref{chic}), 
will be suppressed because of $A^s_0\alt  1$,
the Wilson ratio $R_{\rm W}=1-A^a_0$, 
eq.~(\ref{chis}), will take a value of order unity,
and the system will possess well-defined 
collective modes, namely, zero sounds,
because of $2 \rho^* \Omega f(p,p')\gg 1$.
In other words, the ratio $A/\gamma^2$ 
does not depend sensitively on the Wilson ratio 
as far as $R_{\rm W}\sim 1$.
These characteristics are implicit in ref.~\citen{rf:YY}.
Furthermore, we may say that these systems
are in the regime well away from the Fermi liquid instability,
including the metal-insulator transition:
When $F_1^s\gg 1$ as generally expected from 
$2\rho^* \Omega f(p,p')\gg 1$,
the Drude weight $D$ is not reduced as much 
as expected from the heavily enhanced 
quasiparticle mass $m^*$ since~\cite{rf:I} 
\[ D\propto \frac{1}{m'}=\frac{1}{m^*}\left(
1+\frac{F_1^s}{3}\right)
\gg \frac{1}{m^*}. \]
In general, $D$ will never vanish
as in liquid $^3$He,
for which the `optical mass' $m'=m$ is not renormalized at all
while $m^*$ is heavily enhanced as the pressure is increased.
Still it is remarked that, according to the above definition,
$^3$He cannot be regarded as a stable Fermi liquid
because of $F_0^a\simeq -0.7$ or $A_0^a\simeq -2$.

We saw above that the transport equation (\ref{boltz})
involves the time derivative of $\delta n_p\propto \psi_p$,
and the spatial derivative of $\delta \bar{n}_p\propto  \varphi_p$,
while the collision integral, eq.~(\ref{I(varphi)}),
is written in terms of $\delta \bar{n}_p$
and $A_{pp'k}$.
Therefore, in the static case $\omega=0$,
the effect of the quasiparticle interaction $A_{pp'k}$
does not appear in the transport equation
except in the transition probability, eq.~(\ref{W}).
Then we can discuss quasiparticle transport properties
in general perspective on the assumption of 
the general validity of the Boltzmann equation.

According to the above discussion, 
the universal ratio $A/\gamma^2$ is due to 
the transition probability $W$, eq.~(\ref{W}), 
being proportional to $(\rho^*)^{-2}$,
which we claim is realized in the strong coupling 
Fermi liquids well away from the Fermi liquid instability.
In this case, the energy scale is set solely
by the quasiparticle energy $\varepsilon_p^*$.
For example, the decay rate $\gamma_p^*$ 
is given by
\[\gamma^*_p\simeq \rho^*(\pi T)^2. \]
Thus, 
by the condition
\[|\varepsilon^*_p-\mu|\sim \pi T
\agt \gamma^*_p
\simeq \rho^*(\pi T)^2, \]
the concept of well-defined quasiparticle
is warranted for $T\alt 
T_0\simeq 1/\pi \rho^*$.
This is in contrast to the case
in the vicinity of the Fermi liquid instability,
where $T_0$ can be quite suppressed owing to
an enhanced transition probability $W$.
A general theory based on the Boltzmann equation states
that universal relations must hold 
not only for the resistivity but for 
the other static transport coefficients:
In fact,  as in eq.~(\ref{A/gamma2}),
we can write universal relations for the viscosity $\eta$, 
the thermal conductivity $K$
and the spin-diffusion coefficient $D$ (Appendix).
Here, however,
we show them in a particular case of an isotropic system,
for which we may refer the classical papers.~\cite{rf:AK,rf:Hone}\
For the results of the isotropic system,
we use
\[\gamma=\left(\frac{\pi}{3n}\right)^{2/3}\frac{m^*}{\hbar^2},
\quad n=\frac{k_{\rm F}^3}{3\pi^2},
\]
for $\gamma$ of the specific heat $C=\gamma T$ per particle,
and define the `lattice constant' $a$ by
\[\frac{4\pi}{3} a^3 n=1.\]
Then, under the assumption to neglect 
the momentum dependence of $A_{ pp'k}$,
it is straightforward to obtain the relations
\begin{equation}
KT\gamma^2=\frac{3^{2/3}\pi^{4/3}
}{4^{4/3}\hbar a} 
\alpha^{-1},
\label{KTgam2}
\end{equation}
\begin{equation}
\eta T^2 \gamma^2=\frac{5\hbar 
}{8a^3}
\alpha^{-1},
\label{etaT2gam2}
\end{equation}
\begin{equation}
DT^2 \gamma^3=
\frac{2^{1/3}\pi^{7/3}
a^2}{3^{4/3}\hbar}
\frac{\alpha_{\uparrow\downarrow}^{-1}}{
1-A^a_0}.
\label{DT2gam3}
\end{equation}
The factor $1-A_0^a$ in $D$ 
corresponds to that used in the susceptibility, eq.~(\ref{chis}).
The parameters $\alpha$ and $\alpha_{\uparrow\downarrow}$
are defined in eq.~(\ref{alpha}).
Nevertheless, 
we must note that the averaged quantity $\alpha$
for $\sigma(0)T^2\gamma^2$, eq.~(\ref{A/gamma2}),
and that for $KT\gamma^2$, eq.~(\ref{KTgam2}), for example,
should not be literally identified with each other,
for they are not quite the same 
as their definitions use different types of average of $A_{pp'k}$.
In a lattice system, a term 
due to Umklapp processes is added to $D$ 
unless $\alpha_{\uparrow\uparrow}=0$.
According to the exact solutions,~\cite{rf:Jensen,rf:BrSy} \
eq.~(\ref{KTgam2}) has to be multiplied by a factor $\sim 0.5$,
and eqs.~(\ref{etaT2gam2}) and (\ref{DT2gam3}) by 
$\sim 0.8$.
Therefore, we may write
\begin{eqnarray}
KT \gamma^2&\sim& \frac{0.8}{\hbar a}\alpha^{-1},
\nonumber\\
\eta T^2 \gamma^2 &\sim& \frac{0.5}{a^3}\hbar\alpha^{-1},
\nonumber\\
D T^2 \gamma^3 &\sim& \frac{3
a^2}{\hbar}
\frac{\alpha_{\uparrow\downarrow}^{-1}}{
1-A^a_0}.
\label{UnivRatios}
\end{eqnarray}
These relations 
are approximate and might not be followed well 
by experiment as the relation for 
$A/\gamma^2$ 
is,
since the kinetic coefficients which are mainly determined 
by normal processes may be found to depend sensitively on 
the detailed structure of $A_{pp'k}$ and the Fermi surface.
In this respect also, 
the relation between geometry of the Fermi surface
and the scattering processes relevant to 
kinetic coefficients should be investigated specifically.
It is interesting to note that 
we could estimate not only 
the Lorenz ratio $K/\sigma T$, which may 
take a constant value for a large class of systems
(the Wiedemann-Franz law),
but $A/\gamma^2$ and $KT\gamma^2$ separately.
This is because of the universality 
due to the saturation of
the coupling constant $A^s\simeq 1$
in the strong coupling regime,
where the relevant energy scale is set solely by 
the density of states $\rho^*_k$ of quasiparticle, 
as mentioned above.
Nonetheless, one should still keep in mind the point that
the universality is concluded by
the neglect of the angular dependence of 
the scattering amplitude and 
the detailed structure of the Fermi surface.

\section{Conclusion}
To evaluate a universal value 
taken by the Kadowaki-Woods ratio $A/\gamma^2$ microscopically,
we developed a theory of 
electrical conductivity of Fermi liquids,
following \'Eliashberg~\cite{rf:Eliashberg}
and using the results of Yamada and Yosida.~\cite{rf:YY}\
We derived an equation
to describe quasiparticle transport, by which  
our previous result for the Drude part, eq.~(\ref{Drude}),
and Yamada-Yosida's result, eq.~(\ref{Case;om0k0}),
are reproduced in the coherent and hydrodynamic regime, respectively.   
Moreover, the low-frequency limit of the incoherent part
due to quasiparticle was given in eq.~(\ref{sig1}).
It was shown that some important results of Yamada and Yosida
are interpreted exactly and comprehensively in a general context 
of the phenomenological Landau theory.
So far this point seems not to be pointed out definitely,
presumably because of the thorough use of bare microscopic quantities
in the literature following ref.~\citen{rf:YY}.
In order to write the results just
in terms of renormalized quantities,
assumptions should not be made except those ensuring 
the validity of Fermi liquid concept.
In particular,
one should not make any specific assumption to
derive a formula for the conductivity.
From this point of view,
the Boltzmann equation was obtained as a result.
We consider that
in a microscopic treatment it is important to distinguish
what may be obtained phenomenologically from what is not.
We were mainly concerned with the former.

To clarify the physics underlying the universality 
in the ratio $A/\gamma^2$,  we expressed 
the ratio in a form proportional to 
the product of two factors $\alpha$ and $F$,
and argued that the universal relation $A\propto \gamma^2$ 
is due to constant values taken by these factors.
Moreover, we concluded that 
we must have $\alpha\sim 1$ as well as $F\sim 20$ 
to reproduce the observed universal ratio.
The conjecture for the latter, 
related to a shape of Fermi surface, was borne out 
by evaluating $F$ explicitly in simple examples.
The former, the condition for the 
coupling constant of quasiparticle interaction, was categorized 
as a universality of Fermi liquids in the strong coupling regime.
In contrast with a customary view 
in the heavy fermion problem,~\cite{rf:YY,rf:MMV,rf:KWrel1,rf:KWrel2}\ 
we need not assume heavily enhanced effective mass of quasiparticle,
$z_p^{-1}\gg 1$,
so as to reproduce the relation $A\propto \gamma^2$.
Needless to say, however,
it may happen that the mass enhancement effect and 
the strong coupling condition $\alpha \sim 1$ 
are shown to be correlated to each other in a certain microscopic model.
But we consider that
this point, the universality in that particular model, 
is a secondary problem.
It was claimed for the strong coupling Fermi liquids
that the universal relation in terms of $\gamma$
may be concluded not only for $A$ of resistivity ($A/\gamma^2$=const.),
but for any static kinetic coefficients,
if one may disregard
the effect due to specific geometric features of the Fermi surface 
as compared with the dominant effect due to
the density of states $\rho^*\propto \gamma$.
By way of illustration,
we estimated the ratios \( KT\gamma^2\), \( DT^2\gamma^3 \)
and \( \eta T^2\gamma^2 \) for the thermal conductivity $K$,
the spin-diffusion constant $D$ and the viscosity $\eta$.


\section*{Acknowledgments}
The author would like to thank Professor K.~Yamada
for discussions.
This work is supported  by
Research Fellowships of the Japan Society for the
Promotion of Science for Young Scientists.

\appendix
\section{Transport Coefficients}

As in eq.~(\ref{A/gamma2}),
we can derive
relations for the viscosity $\eta$, 
the thermal conductivity $K$,
and the spin-diffusion coefficient $D$
on the basis of the Boltzmann equation.~\cite{rf:AK,rf:Hone}\
In the following expressions, 
$\alpha$, $\alpha_{\uparrow\uparrow}$
and $\alpha_{\uparrow\downarrow}$ are defined in 
eq.~(\ref{alpha}).
\subsection{Viscosity}
\begin{eqnarray}
\frac{1}{\eta T^2\gamma^2}
&=&
\frac{9}{4\pi}  \alpha F_{\eta},
\end{eqnarray}
where 
\fulltext
\begin{equation}
F_{\eta}=
\frac{\displaystyle
\frac{1}{\Omega^2}\sum_{p,p',k}
\rho_{p}\rho_{p'}\rho_{p-k}\rho_{p'+k}
\mbox{tr} \mib{P}_{p}
(\mib{P}_{p}+\mib{P}_{p'}-\mib{P}_{p'+k}-\mib{P}_{p-k})
}{\displaystyle \rho^4
\left(\frac{1}{\Omega}\sum_p \rho_{p}
\mbox{tr} \mib{P}_{p}\mib{P}_{p}
\right)^2},
\end{equation}\halftext
and
\[\left(\mib{P}_{p}\right)_{ik}\equiv
p_iv_k-\frac{\mib{p\cdot v}}{3}\delta_{ik}.
\]

\subsection{Thermal conductivity}
\begin{eqnarray}
\frac{1}{K T\gamma^2}
&=&\frac{27}{5\pi^4}
\alpha F_{k},
\end{eqnarray}
where
\begin{equation}
F_{k}=
\frac{\displaystyle
\frac{1}{\Omega^2}\sum_{p,p',k}
\rho_{p}\rho_{p'}\rho_{p-k}\rho_{p'+k}
 v_{p\nu}
(v_{p\nu}-v_{p-k\nu})
}{\displaystyle \rho^4
\left(\frac{1}{\Omega}\sum_p \rho_{p}v_{p\nu}v_{p\nu}\right)^2}.
\end{equation}

\subsection{Spin-diffusion coefficient}

\begin{eqnarray}
\frac{1}{DT^2\gamma^3}
&=&
\frac{27}{16\pi^3} (1-A^a_0)
\left(\alpha_{\uparrow\downarrow}F_{d}+
\alpha_{\uparrow\uparrow}F \right),
\end{eqnarray}
where
\fulltext
\begin{equation}
F_{d}=
\frac{\displaystyle
\frac{1}{\Omega^2}\sum_{p,p',k}
\rho_{p}\rho_{p'}\rho_{p-k}\rho_{p'+k}
 v_{p\nu}
(v_{p\nu}+v_{p'\nu}
-v_{p'+k\nu}-v_{p-k\nu})
}{\displaystyle \rho^4
\left(\frac{1}{\Omega}\sum_p \rho_{p}v_{p\nu}v_{p\nu}\right)^2},
\end{equation}\halftext
and $F$ is defined in eq.~(\ref{F}).


\begin{thebibliography}{99}
\bibitem{rf:com0}
  This is necessary to derive eq.~(\ref{zpLambda}).
\bibitem{rf:com}
  At this point, one should not use the Ward identity 
  of the zero temperature formalism
  so as not to 
  have a wrong result with 
  $j^*_{p\mu}$ instead of $v^*_{p\mu}$
  in the right-hand side of eq.~(\ref{zpLambda}).~\cite{rf:YY}\
\bibitem{comment}
In fact, in two dimensional systems, 
the summation in eq.~(\ref{2gamma}) diverges to 
give $\gamma^*_p\propto -T^2\log T$.
Although this anomalous temperature dependence may 
affect some physical quantities,
the `Umklapp' factor 
${\varphi}_{p}+{\varphi}_{p'}-{\varphi}_{p'+k'}-{\varphi}_{p-k'}$
in eq.~(\ref{Case;om0k0}) generally makes the coefficient $A$ of 
$\sigma^{-1}=AT^2$ converge,
with the exception of some singular cases as the system with 
a completely nested Fermi surface.~\cite{rf:FKY,rf:MF}
\bibitem{rf:comment3}
As `Fermi liquid instabilities',
we refer only those instabilities which are
induced when 
macroscopic collective oscillations of the Fermi surface 
becomes unstable.
The instability of microscopic origin,
such as the spin-density wave instability
caused by nesting of the Fermi surface,
does not fall into this category.
\end{thebibliography}
\end{document}